\documentclass[twocolumn]{aastex7}
\usepackage{graphicx} % Required for inserting images
\usepackage{subcaption}
\usepackage{amsmath}
\usepackage{xcolor}

\begin{document}

\title{Characterizing the bolometric-photoevaporative transition in young sub-Neptunes with radiation-hydrodynamic simulations}
\correspondingauthor{William Misener}
\email{wmisener@carnegiescience.edu}

\author[0000-0001-6315-7118]{William Misener}
\affiliation{Earth and Planets Laboratory, Carnegie Institution for Science, 5241 Broad Branch Road NW, Washington, DC 20015, USA}
\email{wmisener@carnegiescience.edu}

\author[0000-0001-6460-0759]{Matthäus Schulik}
\affiliation{Imperial Astrophysics, Department of Physics, Imperial College London,
Prince Consort Road,
London SW7 2AZ, UK}
\affiliation{Department of Earth, Planetary, and Space Sciences, The University of California, Los Angeles,
595 Charles E. Young Drive East,
Los Angeles, CA 90095, USA}
\email{m.schulik@imperial.ac.uk}

\author[0000-0002-0298-8089]{Hilke E. Schlichting}
\affiliation{Department of Earth, Planetary, and Space Sciences, The University of California, Los Angeles,
595 Charles E. Young Drive East,
Los Angeles, CA 90095, USA}
\email{hilke@g.ucla.edu}

\author[0000-0002-4856-7837]{James E. Owen}
\affiliation{Imperial Astrophysics, Department of Physics, Imperial College London,
Prince Consort Road,
London SW7 2AZ, UK}
\affiliation{Department of Earth, Planetary, and Space Sciences, The University of California, Los Angeles,
595 Charles E. Young Drive East,
Los Angeles, CA 90095, USA}
\email{james.owen@imperial.ac.uk}

\begin{abstract}
Hydrodynamic atmospheric escape plays a central role in shaping the demographics of small, close-in exoplanets. Two mechanisms have been proposed to drive mass loss: photoevaporation, powered by UV irradiation, and core-powered mass loss, in which a bolometrically heated wind is sustained by cooling from the planetary interior. Although each mechanism can independently reproduce observed exoplanet demographics, both likely operate simultaneously. To quantify their combined impact, we use AIOLOS, a hydrodynamic radiative transfer code, coupled to a planetary evolution model to self-consistently compute atmospheric escape and planetary evolution. We find that as a typical sub-Neptune contracts, it evolves through distinct escape regimes. The youngest, most inflated planets drive a core-powered, bolometrically heated wind because UV radiation cannot reach the bolometric sonic point. This is followed by a transitional regime shaped by both bolometric and UV heating. As radii decrease further, escape rates approach the purely photoevaporative energy limit. We derive analytic scalings for the transition between these regimes, showing that it occurs at smaller radii for lower-mass and more highly irradiated planets, where core-powered escape dominates. Coupling both processes enhances escape even in more massive, cooler sub-Neptunes. We present the first combined mass-loss rates for a range of planet masses and XUV luminosities and show that the thermal structure below the UV absorption radius—set by atmospheric composition—also affects escape rates. These results integrate core-powered and photoevaporative escape into a unified framework, demonstrating that a self-consistent treatment of atmospheric composition, escape, and evolution is essential for understanding small exoplanets.
\end{abstract} 

\section{Introduction}
A key emerging paradigm in exoplanet science is that atmospheric escape driven by irradiation from the host star is fundamental to the evolution and observed demographics of small ($R \lesssim 4 R_\oplus$), close-in ($P_\mathrm{orb} \lesssim 100$~d) planets. The energy received from their stars is sufficient to strip the hydrogen-dominated primordial envelopes that some small planets accrete from their natal disk over timescales shorter than the gigayears age of typical systems \citep[e.g.][]{LecavelierDesEtangs2007, OwenJackson12, LopezFortney2012,Ginzburg18}. This hypothesis predicts that lower mass planets closer to their host stars experience more vigorous atmospheric escape, neatly explaining the observed `Neptune desert', a paucity of intermediate-sized ($\sim$ 3-10 $R_\oplus$) planets on very short orbital periods \citep[e.g.][]{SzaboKiss2011, MazehHolczer2016}, and the `radius valley', the observed local minimum in the occurrence rate of small planets around Sun-like stars near 1.8 $R_\oplus$ which separates the super-Earth population, with radii less than $\sim 1.8 R_\oplus$, from the sub-Neptune population, with radii between 2 and 4 $R_\oplus$ \citep{Fulton17}. In fact, the radius valley was predicted before its definitive detection by atmospheric escape models \citep{OwenWu13}. 

However, the nature of hydrodynamic escape from small exoplanets is not yet fully resolved, as multiple processes are proposed to act on the small planet population.
First, the loss of pressure support from the disappearing protoplanetary disks leads to a short but important `boil-off' or `spontaneous mass loss' phase lasting $\sim 1$~Myr \citep{OwenWu16, Ginzburg16}, in which cooling and escape are regulated by advection \citep{OwenWu16, Tang2025a} and the disk dispersal timescale \citep{RogersOwen2024}. 
However, since the energy driving escape during the boil-off phase is the release of the binding energy of the contracting envelope itself, this mechanism is unable to fully strip the envelope on its own \citep{Ginzburg16}, requiring long-term hydrodynamic escape. 
Two escape mechanisms have thus emerged to explain the observed exoplanet radius valley and related demographics. The first process hypothesized was `photoevaporation' (PE), in which the outflow is driven by heating from high energy X-ray and UV (XUV) photons from the star \citep[e.g.][]{SekiyaNakazawa1980, lammer2003, Yelle2004, Murray-Clay2009, OwenJackson12}. This heating can produce a rapid outflow, stripping envelopes on short timescales before the high energy stellar radiation declines, after about 100~Myr for Sun-like stars. However, bolometric radiation from the star can drive a strong outflow without any XUV heating. This escaping wind is most vigorous when planets are young and inflated from the energy of formation. Planets naturally cool and contract, decreasing the escape rate, but heat released from the hot, high heat capacity silicate-iron interior can slow contraction, keeping the envelopes inflated and exposed to strong stripping in a process termed `core-powered mass loss' (CPML) \citep{Ginzburg16}.This stripping can, when considered independently from photo-evaporation, take on the order of gigayears \citep[e.g.][]{Gupta20}.

Both mechanisms rely on the same underlying principle: that external heating in the upper atmospheres drives a transonic, hydrodynamic outflow similar to a Parker wind \citep{Parker1958}. Therefore, the broad trends in planet demographics, such as the radius valley and the Neptune desert, are generic to hydrodynamic escape and have been reproduced by both photoevaporative \citep[e.g.][]{OwenWu13, LopezFortney2013, ChenRogers2016, OwenWu17} and core-powered \citep{Ginzburg18, Gupta19, Gupta20} models. These commonalities make these processes difficult to distinguish from demographics alone \citep{RogersGupta2021}. 

To understand for which planets and in what part of their evolution each mechanism dominates, we must delineate the regimes of core-powered mass-loss and photoevaporation. An analytic theoretical framework for the transition between core-powered mass-loss and UV-driven escape was put forward in \citet{OwenSchlichting2024}. The authors found that the key transition between the two mass-loss regimes occurs when XUV photons are able to penetrate the bolometric sonic radius. Initially, when planets are young and have large radii, XUV radiation fails to penetrate the sonic point of the flow, leading to core-powered escape driven by bolometric heating alone. As the planet shrinks, XUV radiation begins to be able to penetrate the sub-sonic region, leading to a regime of enhanced escape that becomes more and more photoevaporation-dominated with further evolution. This hypothesis predicts core-powered escape to dominate the entire evolution of the smallest, highest temperature planets. More massive and cooler planets transition from bolometric to high-energy, EVU/XUV radiation driving their escape, implying that core-powered mass-loss precedes photoevaporation. This is a notable inversion of the predictions inferred from mass loss rates for each mechanism in isolation \citep{OwenWu17,Gupta20}. However, \citet{OwenSchlichting2024} make simplifying assumptions about the temperature structure and evolution of these planets, and as such were unable to fully resolve the details of the critical transition regime between the mechanisms.

Proof-of-concept hydrodynamic simulations verified the notion of regimes where bolometric and XUV heating each dominate \citep{SchulikBooth2023} but did not directly consider the transition from core-powered mass-loss to photoevaporative escape across the sub-Neptune population. While many previous numerical simulations have investigated the onset of photoevaporation \citep[e.g.][]{KubyshkinaFossati2018, AffolterMordasini2023, Tang2025b}, these were limited in their ability to model coupled radiative transfer in the bolometric and UV bands, atmospheric escape rates, and planetary evolution that self-consistently covers all the regimes small planets are expected to pass through. In this work, we overcome these limitations using AIOLOS, a hydrodynamic radiative-transfer code that includes XUV, bolometric, and internal heating \citep{SchulikBooth2023}, coupled to an atmospheric evolution code \citep{MS21, MisenerSchulik2025} to examine sub-Neptune atmospheric escape and self-consistently determine the phases of small planet evolution and the transitions between them.

This self-consistent treatment also allows us to reassess the observational signatures and implications of atmospheric escape models. For example, observations suggest evolution in the ratio of sub-Neptunes to super-Earths takes place on gigayear timescales \citep{BergerHuber2020, DavidContardo2021, ChristiansenZink2023, VachZhou2025}. Based on previous theoretical work indicating that core-powered escape takes place on similar timescales, whereas photoevaporation occurs more quickly \citep{GuptaSchlichting2021}, these observations have been taken as evidence favoring core-powered mechanisms. The difference in instantaneous loss rates has also been used to contextualize the vigor of observed escape from small planets \citep[e.g.][]{ZhangKnutson2023, LoydSchreyer2025}, as atmospheric escape measurements are sensitive to the speed of the outflow \citep{OwenMurray-Clay2023, SchreyerOwen2024}. The mass loss histories of small planets are also vital to understanding the evolution of sub-Neptunes into super-Earths \citep[e.g.][]{MS21}, for instance by constraining fractionation processes \citep{CherubimWordsworth2024, NichollsPierrehumbert2025}. However, these interpretations are based on independent models of each process rather than a comprehensive framework, which only a fully coupled radiative-hydrodynamic-evolution model such as that we present here can provide.

This manuscript is structured as follows. In Section~\ref{sec:escape_defns}, we describe the fundamental physics governing atmospheric escape. In Section~\ref{sec:hydro} we describe our numerical approach using the AIOLOS code to simulate hydrodynamic escape and present our hydrodynamic results on the transition between core-powered and photoevaporative escape. In Section~\ref{sec:evolution}, we implement these escape rates into an evolutionary code and present our findings. We discuss implications and future directions in Section~\ref{sec:discussion} and summarize our conclusions in Section~\ref{sec:conclusion}.

\section{Fundamentals of atmospheric escape}\label{sec:escape_defns}
We begin by introducing key concepts and definitions for the problem of atmospheric escape from small planets. At the most basic level, hydrodynamic escape is caused by incident radiation from a planet's host star. The incident bolometric flux, $F_\mathrm{bol}$, can be quantified in terms of an equilibrium temperature, $T_\mathrm{eq}$, that the outer atmosphere of the planet reaches if it efficiently reradiates the heat: 
\begin{equation}\label{eq:Teq}
    T_\mathrm{eq} = \left(\frac{F_\mathrm{bol}}{4 \sigma}\right)^{1/4} ,
\end{equation}
where $\sigma$ is the Stefan-Boltzmann constant. 

Incident energy from the star heats the atmosphere, causing it to expand in an accelerating wind and unbind gas from the planet. The steady state solution for an isothermal atmosphere is a trans-sonic `Parker-type' wind \citep{Parker1958}. In an atmosphere isothermal at the equilibrium temperature, the mass loss rate is 
\begin{equation}\label{eq:isothermal_loss}
    \dot{M}_\mathrm{iso} = 4 \pi R_\mathrm{s, iso}^2 c_\mathrm{s, iso} \rho_\mathrm{s,iso}
\end{equation}
where the isothermal sound speed $c_\mathrm{s, iso} = (k_\mathrm{B} T_\mathrm{eq}/\mu)^{1/2}$ and $R_\mathrm{s, iso}$ is the isothermal sonic radius, sometimes termed the Bondi radius \citep{OwenSchlichting2024}:
\begin{equation}\label{eq:sonic_radius}
    R_\mathrm{s, iso} = \frac{G M_\mathrm{p}}{2 c_\mathrm{s, iso}^2} = \frac{G M_\mathrm{p} \mu}{2 k_\mathrm{B} T_\mathrm{eq}},
\end{equation}
in which $k_\mathrm{B}$ is the Boltzmann constant and $\mu$ is the mean molecular weight of the atmosphere, $\sim 2$~amu in a molecular hydrogen outflow.

The density at the sonic point, $\rho_\mathrm{s,iso}$, depends on the properties of the deeper atmosphere. We model the deep envelopes of these planets as convective, transitioning to a radiative region at the radiative-convective boundary, $r=R_\mathrm{rcb}$. In the isothermal, hydrostatic limit, the density in the radiative region falls off exponentially:
\begin{equation}\label{eq:exp_isothermal}
    \rho(r) = \rho_\mathrm{rcb} \exp{\left[\frac{2 R_\mathrm{s, iso}}{r} - \frac{2 R_\mathrm{s, iso}} {R_\mathrm{rcb}}\right]}.
\end{equation}
The hydrostatic density profile well-approximates core-powered, isothermal outflows \citep{MisenerSchulik2025}.

Isothermal, bolometrically-driven escape leads to strong outflows from sub-Neptune mass ($M \lesssim 16 M_\oplus$) planets with hydrogen envelopes a few percent of their total mass when the planet has a larger radius, i.e. when $\rho_\mathrm{s, iso}$ is high.
This escape rate then declines with time because small planets contract drastically as they cool \citep[e.g.][]{LopezFortney14}, which chokes the supply of gas exponentially. This contraction can be slowed, however, because the heat capacity of these planets' silicate interiors can surpass that of their atmospheres, and this heat can keep the planets inflated. Maintaining larger radii for longer then drives enough atmospheric escape to explain the radius valley, by the so-called `core-powered mass loss' mechanism \citep[e.g.][]{Ginzburg16, Ginzburg18, Gupta19}. However, we emphasize that this wind is sustained by the incident bolometric heating that is absorbed by the upper envelope of the planet. Hence, core-powered mass-loss relies both on the cooling of the planet's interior and the bolometric stellar radiation. 

In \citet{MisenerSchulik2025}, the authors used hydrodynamical modeling to show that the bolometrically-driven escape from sub-Neptunes depends strongly on the ratio between the atmospheric opacities to incident bolometric opacity $\kappa_{\mathrm{P}, \odot}$ and to outgoing thermal opacities $\kappa_\mathrm{P, therm}$, parameterized as $\gamma$:
\begin{equation}\label{eq:gamma}
    \gamma \equiv \kappa_{\mathrm{P}, \odot}/\kappa_\mathrm{P, therm}.
\end{equation}
This imbalance in the incoming and outgoing radiative transfer leads to the outer envelopes becoming non-isothermal. For $\gamma=1$, i.e. an isothermal atmosphere, \citet{MisenerSchulik2025} found that using an effective $T=T_\mathrm{eq}/2^{1/4}$ in the analytic Parker loss rate instead of $T_\mathrm{eq}$ best matched the numerical models, aligning with previous radiative-transport models \citep{ParmentierGuillot2014}. Other values of $\gamma$ can lead to cool stratospheres or inversions, with $T_\mathrm{out} \propto \gamma^{1/4}$. Similar results were found due to outflows driven by the bolometric irradiation of moons by giant planets \citep{Schulik2025moon}.

However, stars emit not only bolometric luminosity, but also high-energy radiation at UV wavelengths. This emission, the XUV luminosity $L_\mathrm{XUV}$, is observed to decline with time after an early saturation phase \citep[e.g.][]{JacksonDavis2012, TuJohnstone2015, JohnstoneBartel2021}, such that 
\begin{equation}\label{eq:LXUV}
    L_\mathrm{XUV}(t) \approx \begin{cases} 
      L_\mathrm{sat} & t\leq t_\mathrm{sat} \\
      L_\mathrm{sat}(t/t_\mathrm{sat})^{a_0} & t \geq t_\mathrm{sat}
   \end{cases}
\end{equation}
where typical values for Sun-like stars are $L_\mathrm{sat} = 10^{-3.5} L_\odot$, $t_\mathrm{sat} = 100$~Myr, and $a_0=-1.5$ \citep[e.g.][]{JacksonDavis2012}. This UV radiation can heat the upper atmosphere to temperatures much higher than the equilibrium temperature. Since $\dot{M} \propto \rho_\mathrm{s}$ (Eq.~\ref{eq:isothermal_loss}) and the density drops more slowly for hotter gas (Eq.~\ref{eq:exp_isothermal}), these `photoevaporative' outflows are significantly more massive than their bolometrically-driven counterparts. In fact, photoevaporation is often taken to be limited by the total energy intercepted by the planet \citep{WatsonDonahue1981}, such that the energy-limited escape rate $\dot{M}_\mathrm{EL}$ is
\begin{equation}\label{eq:energy_limit_traditional}
    \dot{M}_\mathrm{EL} = \eta F_\mathrm{XUV} \frac{\pi R_\mathrm{XUV}^2 R_\mathrm{t}}{G M_\mathrm{p}}
\end{equation}
where $\eta$ is an efficiency factor usually on the order of 0.1, $F_\mathrm{XUV}$ is the XUV flux received by the planet. This equation depends on two key radii in these planets' atmospheres. The transit radius, $R_\mathrm{t}$, is the radius at which the chord optical depth to incident stellar radiation is equal to 1, i.e.:
    \begin{equation}\label{eq:transit_integral}
        1 = 2 \int_{R_\mathrm{t}}^{\infty} \rho(r) \kappa_{\mathrm{R}, \odot} \mathrm{d}s
    \end{equation}
where $s = (r^2 - R_\mathrm{t}^2)^{1/2}$ is the chord length and $\kappa_{\mathrm{R}, \odot}$ is the atmospheric Rosseland mean opacity to stellar radiation. This radius typically occurs at a pressure  $\sim 20$~mbar, about 10 scale heights larger than the radiative-convective boundary $R_\mathrm{rcb}$ \citep[e.g.][]{LopezFortney14, RogersOwen2024, TangFortney2024}, which was neglected in some past studies of escape from young planets \citep[e.g.][]{MS21}.

Meanwhile, $R_\mathrm{XUV}$ is the radius at which the XUV radiation is absorbed. For compact atmospheres and/or very massive planets, all radiation is absorbed at nearly the same radius, such that $R_\mathrm{t} \sim R_\mathrm{XUV}$. But for low mass, inflated planets, XUV can be absorbed far above $R_\mathrm{t}$ \citep[e.g.][]{KubyshkinaFossati2018, BroomeMurray-Clay2025}. Our hydrodynamic radiative-transfer models self-consistently account for this effect; however, it is useful to roughly approximate the XUV radius analytically by extending an isothermal atmosphere from the transit radius using Eq.~\ref{eq:exp_isothermal}:
\begin{equation}\label{eq:RXUV}
    R_\mathrm{XUV} \approx \cfrac{R_\mathrm{t}}{1+\cfrac{R_\mathrm{t}}{2 R_\mathrm{s, iso}} \ln{\left(\cfrac{P_\mathrm{t}}{P_\mathrm{XUV}}\right)}},
\end{equation}
where $P_\mathrm{t} \sim 20$~mbar is the characteristic pressure at the transit radius \citep[e.g.][]{LopezFortney14} and $P_\mathrm{XUV} \sim 1$~nbar is the characteristic pressure at the XUV absorption radius \citep[e.g][]{Murray-Clay2009, KrennFossati2021}.

\begin{figure*}[ht!]
    \includegraphics[width=\textwidth]{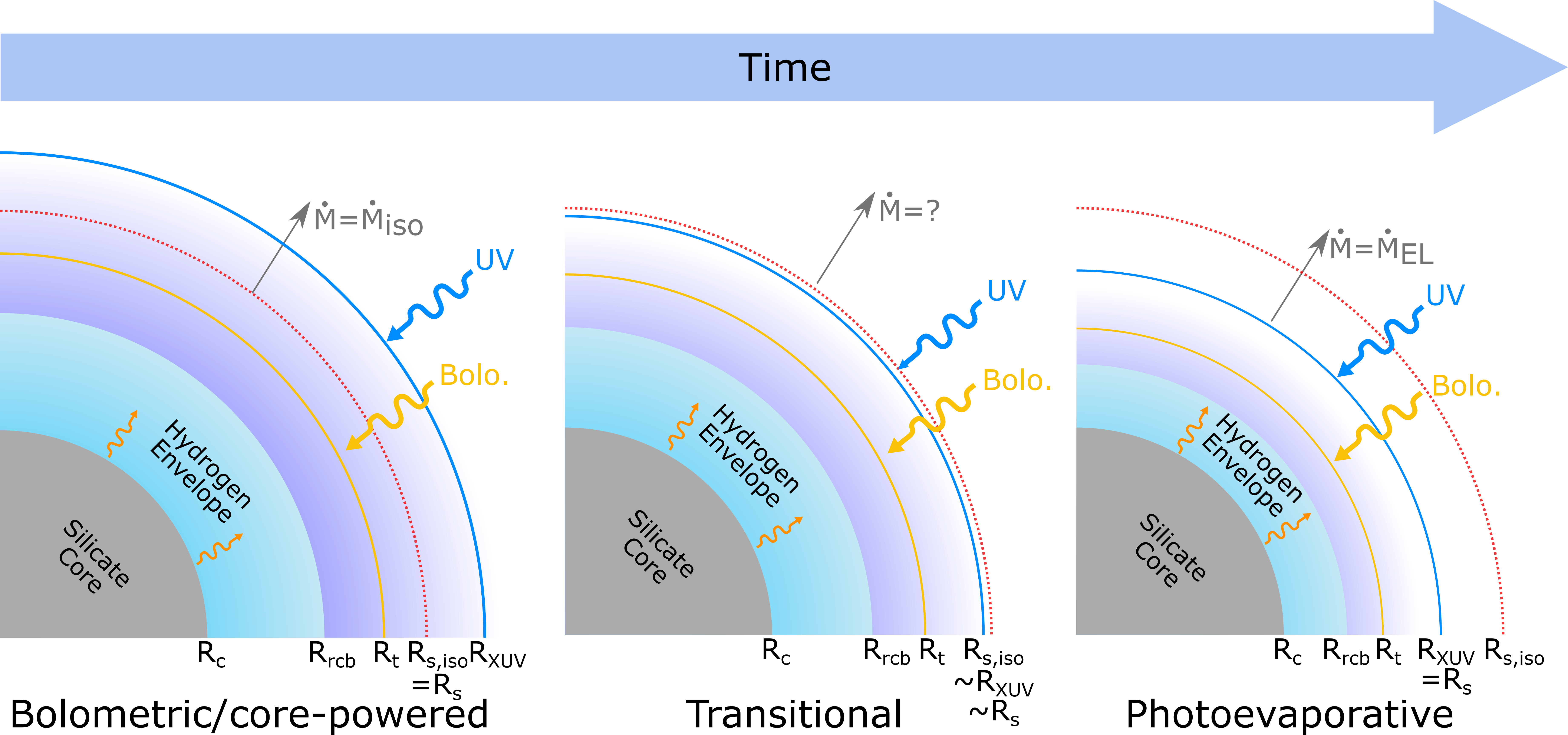}
    \caption{Diagram (not-to-scale) illustrating the key mass loss regimes we investigate in this work and the critical radii they depend on. Planet age increases left to right, while planet size decreases. XUV radiation penetrates to a depth of $R_\mathrm{XUV}$ (blue). For the largest planet, on the left, this high-energy radiation is unable to penetrate the sonic radius of an isothermal outflow at the equilibrium temperature, $R_\mathrm{s, iso}$ (red), and so is unable to influence the mass loss. This wind is fully bolometrically powered, and heat from the silicate-iron core can help sustain the planet's radius at such a size, driving core-powered mass loss. Conversely, for the smallest planet, on the right, this radiation penetrates well within the sonic radius, leading to an energy-limited photoevaporative wind. In the transitional case, in the middle, the UV is just able to penetrate the sonic radius, at which point the outflow is already nearly supersonic. The exact physics of this intermediate regime, and thus the resultant mass loss rate, is not clear from analytic approaches and requires a hydrodynamic model to resolve. Figure partially adapted from \citet{OwenSchlichting2024}.} 
    \label{fig:diagram}
\end{figure*}

The various radii defined in this section and their relation to each other in the different mass loss regimes are shown in Figure~\ref{fig:diagram}. \citet{OwenSchlichting2024} used analytic formulations like those above to find that the relative importance of XUV- and bolometrically-driven escape depends on two factors. The first is whether the XUV heating is absorbed within the isothermal sonic point, which they term the Bondi radius, i.e. comparing $R_\mathrm{s, iso}$ (Eq.~\ref{eq:sonic_radius}) and $R_\mathrm{XUV}$ (Eq.~\ref{eq:RXUV}).  This transition is depicted in Fig.~\ref{fig:diagram}: on the right, for the planet with the largest radius, $R_\mathrm{XUV} > R_\mathrm{s,iso}$, meaning the mass loss rate is set by the bolometric heating. For the smallest radius planet on the left, UV radiation is able to penetrate deeply, driving energy-limited photoevaporative escape. \citet{OwenSchlichting2024} found that this transition occurs at $R_\mathrm{t}/R_\mathrm{s, iso} \approx 0.11$, a constant value (their Eq.~13), and that this transition always occurs before the bolometrically-heated outflow becomes collisionless. However, the escape behavior in the intermediate transition region between the two endmembers was unresolved in their analytic approach. The second factor is whether the XUV heating produces sufficient energy to drive a stronger wind, or in other words, whether the UV heating is sufficient to increase the temperature beyond the equilibrium temperature.
In this case, \citet{OwenSchlichting2024} found that the transition is to first order constant in $R_\mathrm{t}/R_\mathrm{s,iso}$, though it depends logarithmically on $F_\mathrm{XUV}$ (their Eq.~23). We test these findings, and therefore elucidate the transition between core-powered and photoevaporative escape, using radiation hydrodynamic simulations that self-consistently include both bolometric and XUV radiative transfer via the methods described below.

\section{Hydrodynamic Modeling}\label{sec:hydro}
In this section, we describe the self-consistent hydrodynamic radiative transfer modeling we undertake in this work. We begin with our methods in Sec.~\ref{sec:hydromethods} then present our atmospheric structure and escape rate findings in Sec.~\ref{sec:hydroresults}.

\subsection{Hydrodynamic Methods}\label{sec:hydromethods}
In this work, we use AIOLOS, a 1D radiation-hydrodynamics code \citep{SchulikBooth2023}\footnote{Publicly available at \url{https://github.com/Schulik/aiolos}, code used for this work available on zenodo at \citet{aiolos_zenodo}}, which has been used to investigate escape in a variety of planetary environments \citep{BoothOwen2023, Schulik2025moon, Schulik2025helium}. We use the methods developed in \citet{MisenerSchulik2025}, which we summarize below, with adaptations to allow for photoevaporative escape. We connect these simulations to analytic structure models, which well-approximate the deeper regions of the envelope where we expect convection and radiative diffusion to dominate, to link AIOLOS outputs to planetary characteristics such as atmospheric mass and cooling timescales.

\subsubsection{AIOLOS setup and run parameters}
While we briefly describe our simulation set up here, we refer the reader to \citet{SchulikBooth2023} and \citet{MisenerSchulik2025} for more details. AIOLOS uses a HLLC Riemann solver \citep{toro1994} to solve a per-species Euler system \citep{toro1994}, utilizing the well-balancing scheme of \citet{KappeliMishra2016}.

A major change from the AIOLOS simulations of \citet{MisenerSchulik2025} is the inclusion of four species in the hydrodynamic outflow: molecular hydrogen (H$_2$), neutral atomic hydrogen (H$^0$), protons (p+), and electrons (e-). The inclusion of multiple species was not necessary in \citet{MisenerSchulik2025} due to its restriction to low temperature, bolometrically-driven outflows, which were not hot enough to dissociate H$_2$. However, hydrogen readily dissociates under the UV instellations we consider in this work, and these changes in mean molecular weight and speciation play crucial roles in the outflow physics. We include thermal and UV dissociation and ionization reactions. Recombination coefficients and reaction rates are taken from \citet{Yelle2004}, we show those used in Table~\ref{tab:rxns}.
\begin{table}
    \centering
    \begin{tabular}{cccc}
        Reaction & $\alpha$ & $\beta$ & $\gamma$\\ \hline
        1 e- + 1 p+ $\rightarrow$ 1 H$^0$ & $1.074889 \times 10^{-9}$ & -0.9 & 0. \\
        2 H$^0$ + M $\rightarrow$ 1 H$_2$ + M & $2.45\times 10^{-31}$ & -0.6 & 0.\\
        1 H$_2$ + M $\rightarrow$ 2 H$^0$ + M & $1.5 \times 10^{-9}$ & 0.0 & $4.8\times 10^{4}$\\
    \end{tabular}
    \caption{Arrhenius-type reaction coefficients for our thermal reaction network. The reaction rates are $r=\alpha \times T^{\beta}\times \exp(-\gamma/T)$ with $\mathrm{d}n_s/\mathrm{d}t = -r n_\mathrm{A} n_\mathrm{B}$.}
    \label{tab:rxns}
\end{table}
Further reactions are omitted from our model to preserve physical clarity. Ionization and dissociation cross sections are from PHIDRATES \citep{huebnermukherjee2015}. For an XUV luminosity of $10^{-3.5} L_\odot$, equivalent to a high-energy flux of $6 \times 10^4$~erg cm$^{-2}$ s$^{-1}$ at a semi-major axis corresponding to $T_\mathrm{eq} = 1000$~K, we find an optically thin photoionization rate of $3.11 \times 10^{-4}$~s$^{-1}$ and an optically thin photodissociation rate of $4.60 \times 10^{-3}$~s$^{-1}$.

We use four radiation bands: three representing incoming radiation and one representing thermal outgoing radiation. The three incoming radiation bands represent low energy stellar photons (wavelengths $\lambda > 110$~nm, $E<11.2$~eV), those with sufficient energy to dissociate H$_2$ but not ionize H (91~nm $< \lambda <$ 110~nm, $11.2$~eV $< E < 13.6$~eV), and those with sufficient energy to both dissociate H$_2$ and ionize H ($\lambda < 91$~nm, $E>13.6$~eV). Each band can have independent opacities, which we discuss in Section~\ref{sec:opacities}.
In this work, we vary the incident bolometric and high-energy fluxes as independent variables. We use high-energy luminosities between $10^{-5}$ and $10^{-3}$ times the bolometric luminosity of a Sun-like host star, encompassing the range of typical exoplanet host stars \citep{JacksonDavis2012}, as well as $10^{-13}$ times the bolometric luminosity to represent a UV-free case.

As in \citet{MisenerSchulik2025}, we vary the inner boundary of the AIOLOS simulations to correspond to different planet sizes. The inner boundary of the simulation is held at a fixed density of $\rho=10^{-5}$~g\ cm$^{-3}$ ($P\approx 0.4$~bar at 1000~K), which corresponds to an optical depth of order $\sim 10-100$ for the single-temperature Rosseland mean opacities we use (see Eq.~\ref{eq:Rosseland_scale} below), but it is allowed to vary in temperature. We use a variable grid-spacing, 500 cells per decade in an inner region ($r < 8 \times 10^9$~cm) to resolve shocks that might occur during the initial simulation burn-in phase, before a smooth hydrodynamic outflow is established, 
and 200 cells per decade in the outer region. We choose the outer open boundary location on a per-run basis to include the sonic point and UV absorption radius, balancing with computation-time considerations. Its radius is typically of order $10^{11}$~cm in the runs presented here. The profiles are initialized with a hydrostatic density profile isothermal at the planetary equilibrium temperature with an H$_2$-dominated composition ($n_\mathrm{H^0}/n_\mathrm{H_2} = 10^{-8}$), and the UV flux ramps up over 10$^3$~s of simulation time. We use a Courant number of 0.1 \citep{Courant1928, toro2009}. Each simulation typically takes $\sim 10^7$~s of simulation time to reach steady-state, such that the mass loss rate is changing by $\lesssim 1$ percent over $10^6$~s.

\subsubsection{Opacities}\label{sec:opacities}
Following \citet{MisenerSchulik2025}, we use simplified atmospheric opacities to better understand the basic physics at hand. High up in the atmosphere, Planck mean opacities can describe the thermal properties. In our fiducial model, we assume the Planck mean opacities to incident bolometric (i.e. two-temperature) and outgoing thermal (i.e. single-temperature) radiation to be equal: $\kappa_\mathrm{P, \odot} = \kappa_\mathrm{P, therm} = 7.5$~cm$^2$~g$^{-1}$ (i.e. $\gamma=1$ as defined in Eq.~\ref{eq:gamma}). Previous work found that bolometrically-driven atmospheric escape mechanisms can be sensitive to this opacity ratio \citep{MisenerSchulik2025}. To test whether this sensitivity extends to the transition between bolometric and photoevaporative escape, we test atmospheric $\gamma$ values of 0.1 and 10, which we achieve by varying $\kappa_\mathrm{P, \odot}$ down and up by a factor of 10, respectively. Increasing $\kappa_\mathrm{P, \odot}$ is physically analogous to, e.g., increasing the aerosol content of the atmosphere, since aerosols are good visible absorbers, or to increasing the stellar effective temperature, thereby shifting the stellar blackbody toward visible regions with strong absorbing lines \citep[e.g.][]{Freedman14}. Alternatively, altering $\kappa_\mathrm{P, therm}$ to achieve the same changes in $\gamma$ is physically analogous to changing the metallicity of the atmosphere, since heavy molecules tend to be good infrared absorbers. In Appendix~\ref{sec:appendix_opacity}, we test our opacity assumptions and find that our results are relatively insensitive both to the absolute values of $\kappa_\mathrm{P}$ and to which opacity is varied: to first order the value of $\gamma$ determines the escape behavior. We take the opacity to ultraviolet radiation $\kappa_\mathrm{P, UV} = 10^6$~cm$^2$~g$^{-1}$. We neglect the intra-band wavelength dependence of these opacities, which can lead to changes to the thermal profiles and thus the escape rates if large opacity windows are present \citep{Schulik2025moon}, as the additional numerical complexity is computationally prohibitive for the scope of this work.

Deeper down, when the optical depth $\tau > 1$ to outgoing radiation, the thermal profile is determined by radiative diffusion and characterized by Rosseland mean opacities. Single-temperature Rosseland mean opacities, $\kappa_\mathrm{R, therm}$, scale strongly with density, while Planck means are relatively insensitive \citep[e.g.][]{Freedman14}, so we adopt a density scaling such that
\begin{equation}\label{eq:Rosseland_scale}
    \kappa_\mathrm{R, therm}=0.1 \left(\frac{\rho}{10^{-3}\mathrm{\,g\,cm^{-3}}}\right)^{0.6}\mathrm{\,cm^2\,g^{-1}},
\end{equation}
following e.g. \citet{Gupta19}, \citet{MS21}, and \citet{MisenerSchulik2025}. Since the transit radius also probes the $\tau \sim 1$ surface, to calculate it in Eq.~\ref{eq:transit_integral} we use a solar Rosseland opacity $\kappa_\mathrm{R, \odot} = 0.001$~cm$^2$~g$^{-1}$ \citep{Freedman14, OwenSchlichting2024, TangFortney2024}. This approach is an improvement on \citet{MisenerSchulik2025}, which used the solar Planck opacity for this calculation and thus tended to overestimate the transit radii, as Planck means, sensitive to opacity maxima, are higher than Rosseland means, sensitive to opacity minima. Since the transit radius is a white light measurement, the Rosseland mean is a more appropriate approximation.

\subsection{Radiation-Hydrodynamic Simulation Results}\label{sec:hydroresults}
\subsubsection{A transition from bolometric- to UV-heated outflows}
\begin{figure*}
\hspace*{-0.5cm}
\begin{subfigure}{0.5\textwidth} 
   \centering
   \includegraphics[width=\textwidth]{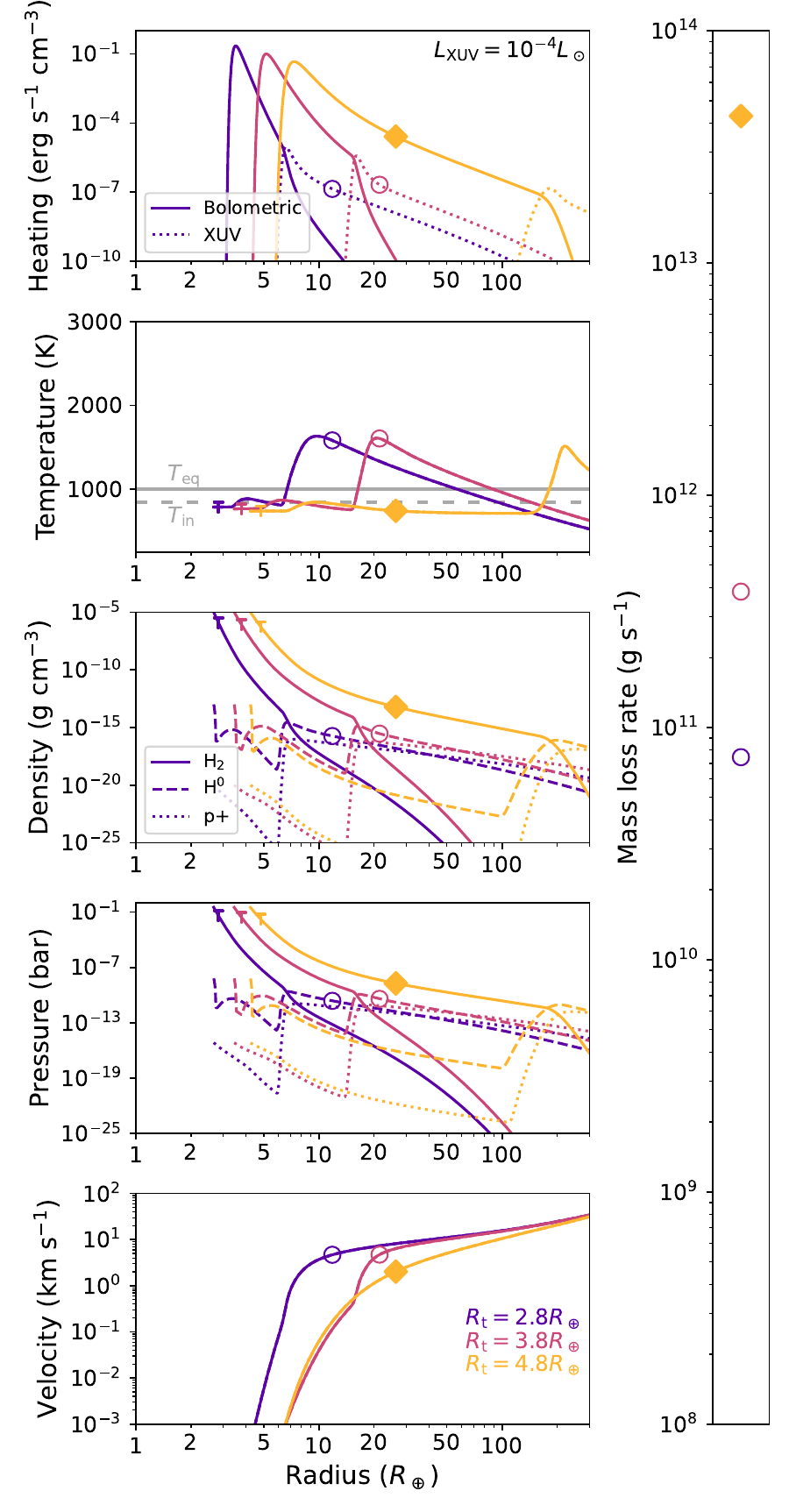}
\end{subfigure}
\begin{subfigure}{0.5\textwidth} 
   \centering
   \includegraphics[width=\textwidth]{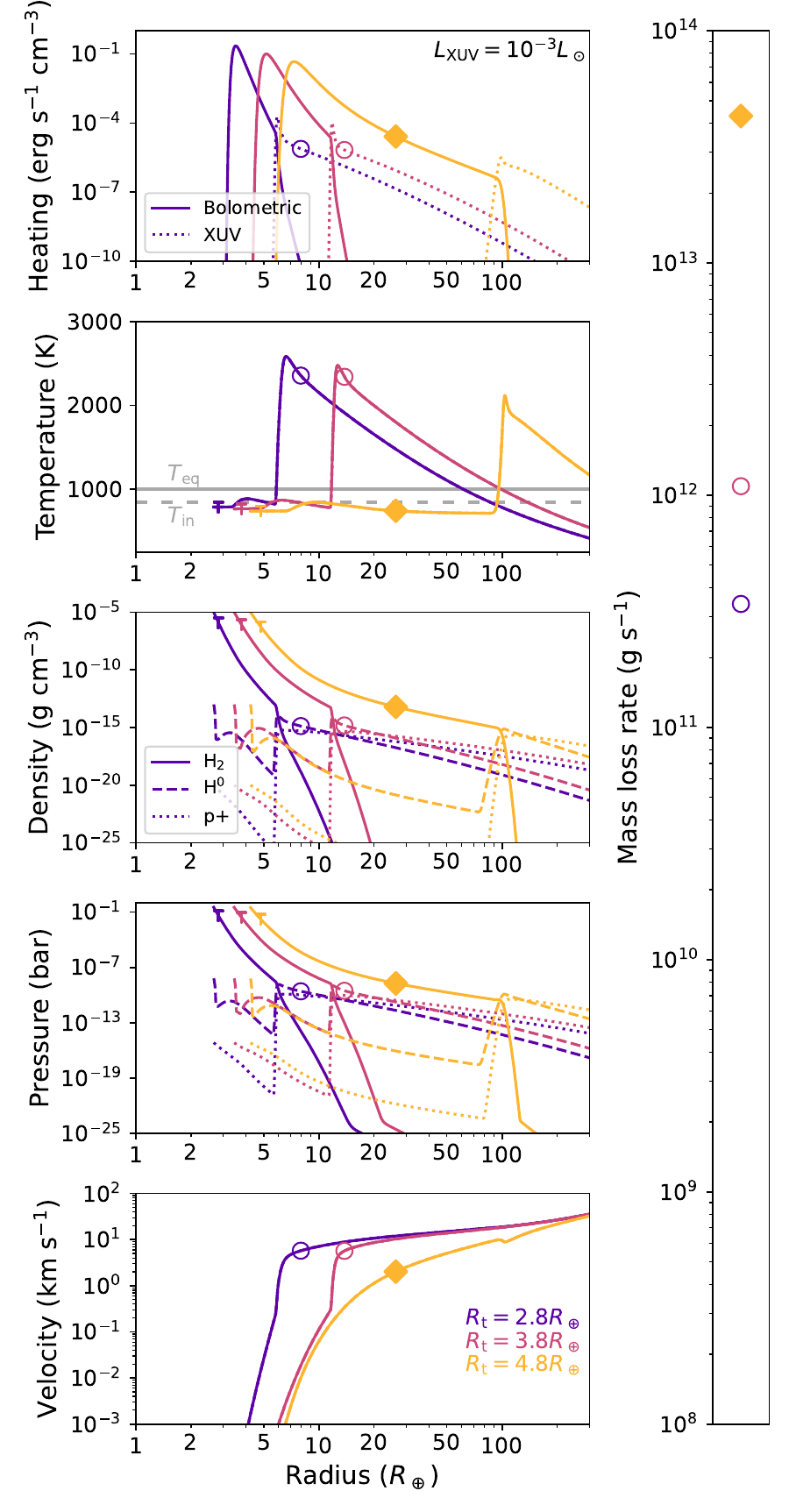}
\end{subfigure}
\caption{Atmospheric radial profiles of (top-to-bottom) the local heating rate, temperature, density, pressure, and velocity for a planet with $\gamma = 1$, $M_\mathrm{p} = 5 M_\oplus$, and $T_\mathrm{eq} = 1000$~K. Left is $L_\mathrm{XUV}/L_\mathrm{bol} = 10^{-4}$, right is $L_\mathrm{XUV}/L_\mathrm{bol} = 10^{-3}$. Colors represent different transit radii (marked with a T). A symbol denotes the sonic radius of the outflow: a thick diamond if the flow is H$_2$-dominated, and a thin diamond if the flow is H$^0$-dominated. The corresponding mass loss rates are shown along the bar to the right. Heating by XUV radiation penetrates the sonic radius for $R_\mathrm{t} \lesssim 4.5 R_\oplus$ (the darker-colored 2.8 and 3.8 $R_\oplus$ runs), increasing the mass loss rate above the core-powered escape rate. Meanwhile, at large radii, the XUV radiation fails to penetrate the flow, as shown by the lighter-colored, 4.8 $R_\oplus$ run, leading to a bolometrically-driven wind. The larger XUV flux on the right leads to hotter temperatures in the UV portion of the flow and larger escape rates for the same planet radius for the simulations with UV-driven winds, but no change for the bolometrically-driven winds.}
\label{fig:Rcomp_fixedUV}
\end{figure*}
In Figure~\ref{fig:Rcomp_fixedUV}, we compare the radial profiles in local heating rate, temperature, density, pressure, and velocity and the corresponding mass loss rates for simulations of a $M_\mathrm{p} = 5 M_\oplus$, $T_\mathrm{eq} = 1000$~K (equivalent to semi-major axis $0.078$~au and orbital period $7.9$~days around a Sun-like star) planet with $\gamma = 1$. The left and right panels show results for a stellar UV luminosity $10^{-4}$ and $10^{-3}$ times the bolometric luminosity, equivalent to incident UV fluxes of $2.4 \times 10^4$ and $2.4 \times 10^5$~erg~cm$^{-2}$~s$^{-1}$, respectively. The colors correspond to planetary transit radii, with lighter colors representing larger-radius planets. In the heating profiles, bolometric heating is represented by a solid line, while the dotted line traces UV heating. In the density and partial pressure profiles, the different line styles denote the profiles of molecular, atomic, and ionized hydrogen. A `T' symbol marks the transit radius (typically at $P \sim 10$~mbar, e.g. \citealt{LopezFortney14}), while a circle or diamond denotes the sonic radius of the outflow, depending on whether the outflow is atomic or molecular at the sonic radius, respectively. The corresponding mass loss rates for each simulation are shown to the right.

In both UV luminosity cases, for the largest planet shown (light orange, $R_\mathrm{t} = 4.8 R_\oplus$), the UV absorption radius is well outside the sonic radius of the bolometrically heated, $T = T_\mathrm{in} \sim 800$~K outflow, at $\sim 26 R_\oplus$. Therefore, the UV heating does not affect the outflow at all: the density profile of the bolometrically heated region fully controls the hydrodynamic wind. The mass loss rate is quite high due to the large size of the planet, with $\dot{M} \approx 10^{14}$\ g~s$^{-1}$, but it is insensitive to the incident UV flux, as shown by the orange mass loss rates being equal in the left and right of Figure~\ref{fig:Rcomp_fixedUV}. The outflow remains molecular into the supersonic region. A stationary shock appears to form at the supersonic UV absorption front in these runs, near $100 R_\oplus$ in the $10^{-3} L_\odot$ case; we discuss this behavior further in Appendix~\ref{sec:shock}.

For the smaller-radius planets, shown by darker lines, incident UV light penetrates below the bolometric sonic radius, causing a factor of 2-3 spike in temperature and photochemically dissociating or ionizing much of the molecular hydrogen. This heating accelerates the flow, which goes supersonic shortly outside of the UV absorption radius, which is typically at $\rho \sim 10^{-14}$\ g~cm$^{-3}$ and $P \sim 10^{-10}$~bar. At the sonic point for these smaller-radius planets, the outflow is dominated by atomic neutral hydrogen, with protons increasing in number in the supersonic region.

\begin{figure}
\hspace*{-0.5cm}
\centering
\includegraphics[width=\columnwidth]{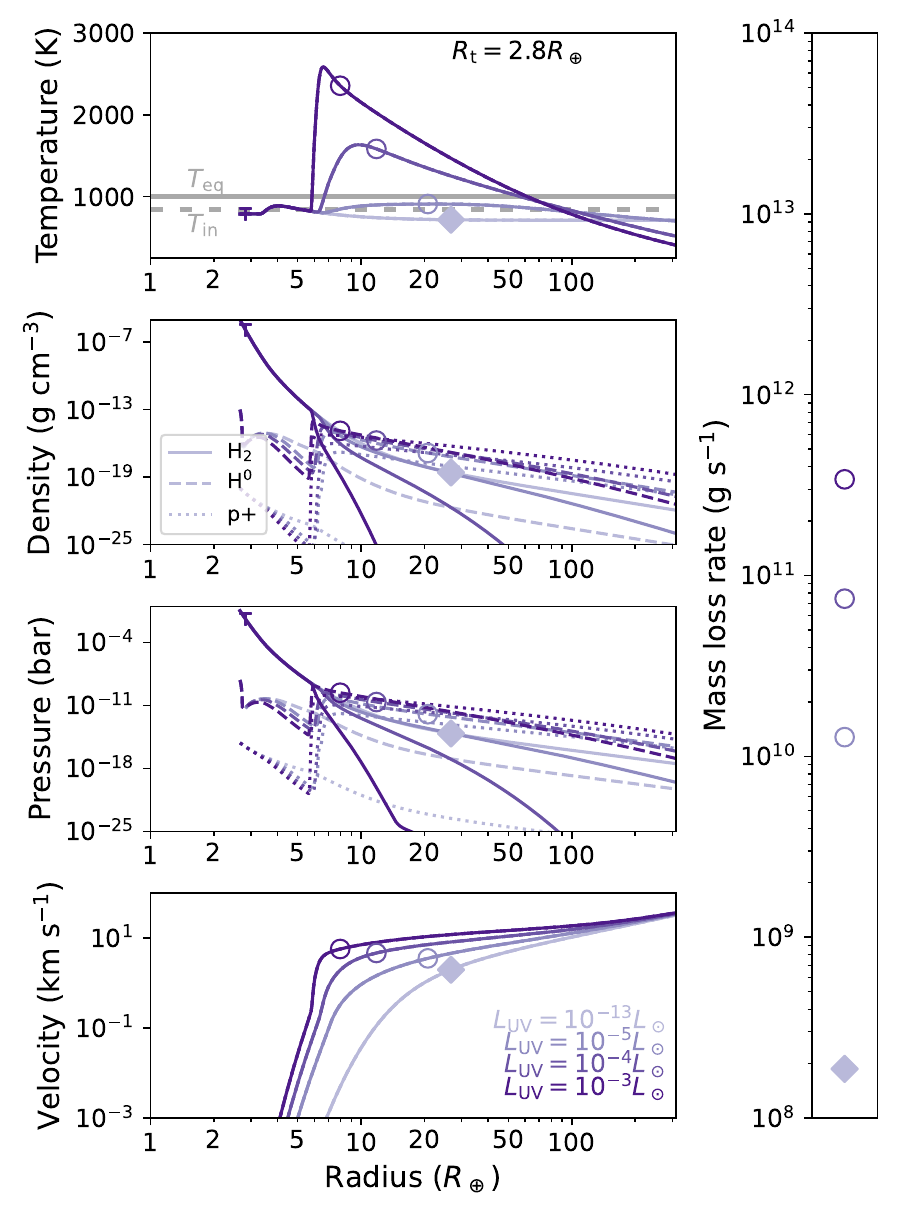}
\caption{Atmospheric profiles for $\gamma \sim 1$, $M_\mathrm{p} = 5 M_\oplus$, $T_\mathrm{eq} = 1000$~K, and $R_\mathrm{t} = 2.8 R_\oplus$. Shading represents different incident XUV fluxes of $10^{-13} (\sim 0), 10^{-5}, 10^{-4},$ and $10^{-3} \times$ the bolometric flux. A diamond denotes the sonic radius of the outflow. In this case, XUV penetrates below the sonic radius, increasing loss rates. The loss rate scales with the incident energy, but much of the density profile is set by the bolometric heating.}
\label{fig:UVcomp_fixedR}
\end{figure}

For the lower flux, $L_\mathrm{XUV} = 10^{-4}$ case on the left of Figure~\ref{fig:Rcomp_fixedUV}, the heating due to XUV radiation is weaker, leading to lower temperatures. This weaker heating leads to weaker acceleration in the flow: the sonic point is reached further out and at lower densities for the lower flux case. This effect is shown more clearly in Figure~\ref{fig:UVcomp_fixedR}, which compares profiles generated with fixed planet parameters and radii (those of the purple curves in Figure~\ref{fig:Rcomp_fixedUV}) but changing the stellar UV luminosity. The luminosities increase from a negligible value (lightest shade) to 10$^{-3} L_\odot$ (darkest), leading to increasing heating and thus a smaller radius over which the flow is accelerated to supersonic speeds. The heating also increases the mass loss rate, although slightly more weakly than the linear trend with UV flux predicted by the energy limit (Eq.~\ref{eq:energy_limit_traditional}), which is due to efficient molecular hydrogen cooling \citep{SchulikBooth2023}. The bolometric outflow sets a floor to the thermal profile and thus the escape rate at low UV luminosities, corresponding to the heating limit discussed in \citet{OwenSchlichting2024}.

\begin{figure*}
\centering
\begin{subfigure}{\columnwidth} 
    \includegraphics[width=\textwidth]{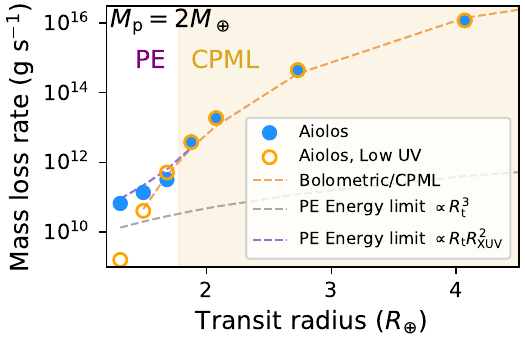}
\end{subfigure}
\\
\begin{subfigure}{\columnwidth} 
    \includegraphics[width=\textwidth]{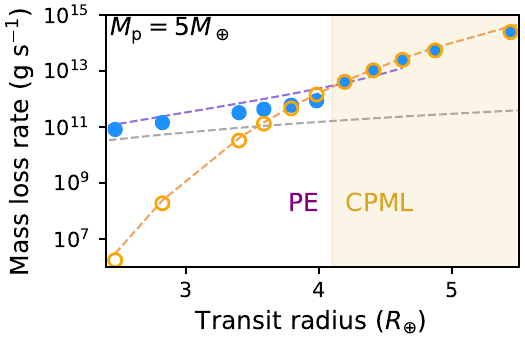}
\end{subfigure}
\\
\begin{subfigure}{\columnwidth} 
    \includegraphics[width=\textwidth]{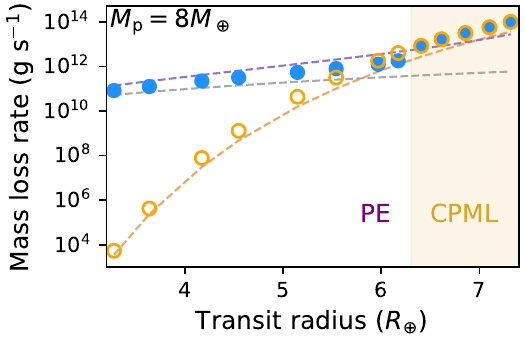}
\end{subfigure}
\caption{Mass loss rates as functions of transit radii for three planet masses of 2, 5, and 8 $M_\oplus$. Filled blue circles are the results of AIOLOS hydrodynamic radiative-transfer simulations for $T_\mathrm{eq}=1000$~K, and $L_\mathrm{UV}/L_\mathrm{bol} = 10^{-3.5}$. Unfilled yellow circles are the results of AIOLOS runs with the same parameters but with negligible $L_\mathrm{UV}$. Shown by a yellow dashed line is the isothermal Parker limit calculated using $T=T_\mathrm{eq}/2^{1/4}$ (Eq.~\ref{eq:isothermal_loss}), shown in \citet{MisenerSchulik2025} to be a good approximation for core-powered, bolometrically-driven escape. The gray and purple dashed lines show analytic energy limited escape rates using Eq.~\ref{eq:energy_limit_traditional}, with an efficiency $\eta=0.1$. The gray line approximates $R_\mathrm{XUV} \sim R_\mathrm{t}$, while that in purple analytically approximates $R_\mathrm{XUV}$ per Eq.~\ref{eq:RXUV}. At high $r$, the mass loss rates are very close to the core-powered, bolometrically heated rate, indicating the flow is unaffected by UV absorption at these radii and in the core-powered regime (shaded yellow and labeled CPML). As the radius decreases, the mass loss rates enter a transitional regime in between the core-powered and photoevaporative rates, converging on the energy-limited rate at low $r$. Using $R_\mathrm{XUV} > R_\mathrm{t}$ is vital to correctly approximating the photoevaporative escape rate.}
\label{fig:Mdot_comparison}
\end{figure*}

We compare a suite of simulation results against analytic mass loss rates for the bolometrically-heated and energy-limited photoevaporative cases in Figure~\ref{fig:Mdot_comparison}. 
We find that at large radii, the outflow is well-characterized by an isothermal Parker wind and is not sensitive to the UV irradiation. Hence, the simulated escape rates for the UV case (blue dots) are coincident with both the no-UV simulations (yellow circles) and the bolometrically heated, isothermal approximation (yellow dashed line). At some radius, below about $4.2~R_\oplus$ for the $5 M_\oplus$ case, UV can penetrate the flow below the isothermal sonic radius. We term this location the transition radius, at which point the planets transition to a flow shaped by photoevaporation. For the same planet mass but smaller transit radius, the UV penetrates more deeply, leading to a strong enhancement in escape rate over a planet receiving no UV irradiation (see the dark curves in Fig.~\ref{fig:UVcomp_fixedR}). The mass loss rate typically reaches a value in between the photoevaporative energy limit and the bolometric limit. Towards the smallest radii, less and less of the density profile is set by the bolometric temperature, and the atmospheric escape rate approaches the energy limit. These radiation-hydrodynamic simulation results are therefore consistent with the escape regimes put forward in \citet{OwenSchlichting2024}. We also find that the transition radius increases with planet mass, a trend we will examine further in Sec.~\ref{sec:results_trends}.

While the resultant mass loss rates show a transition between the bolometric and energy-limited approximations, we find additional complexity not captured by the simple analytic approximations often taken in the literature. Firstly, care must be taken in analytically approximating the photoevaporative energy limit, as we observe mass loss rates exceeding those calculated using Eq.~\ref{eq:energy_limit_traditional} using $R_\mathrm{XUV} \sim R_\mathrm{p}$. This enhancement in escape rates over the naive energy limit is unsurprising and in line with previous works \citep[e.g.][]{KubyshkinaFossati2018, OwenSchlichting2024, BroomeMurray-Clay2025}, as the UV heating occurs well above the planetary transit radius, as shown in Fig.~\ref{fig:Rcomp_fixedUV}. We find that approximating $R_\mathrm{XUV}$ via Eq.~\ref{eq:RXUV} better captures our simulation escape rate trends, and that $\eta \sim 0.1$ is an appropriate efficiency value in this regime. However, our results reveal additional hydrodynamic complexity not captured by, e.g., taking the maximum of two analytic approximations: at radii just below the transition radius, for all three planet masses shown here, the mass loss rate from our hydrodynamic simulations decreases below both analytic predictions as well as the zero-UV flux escape rate. In this regime, the UV radiation only barely penetrates below the bolometric sonic radius, $R_\mathrm{s, iso}$, so the densities have declined exponentially, following the scale height set by the equilibrium temperature. The flow heats up and dissociates where the UV is absorbed, increasing the sound speed but decreasing the sonic radius per Eq.~\ref{eq:sonic_radius} such that the outflow is nearly immediately accelerated to the sound speed. However, the UV heating is unable to raise the sonic density much above its value in an equilibrium outflow. Therefore, to understand the decrease in escape rate, we can compare the terms in Eq.~\ref{eq:isothermal_loss}: while the sound speed is higher in the UV-heated outflow compared to a bolometric outflow, the sonic density is nearly the same, and the sonic radius is lower, leading to lower escape rates just inside the transition radius.
 
\begin{figure*}
\centering
\begin{subfigure}{\columnwidth} 
    \includegraphics[width=\textwidth]{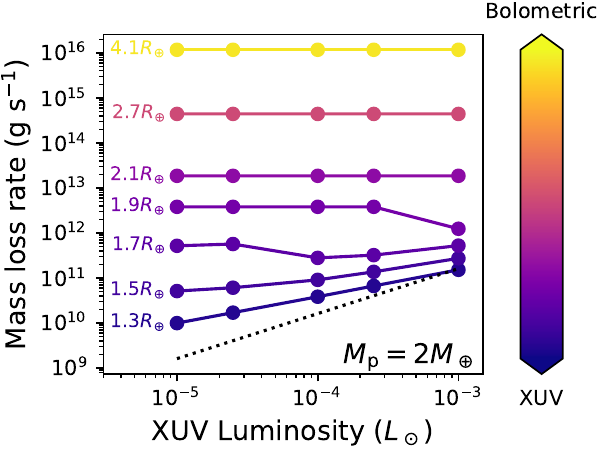}
\end{subfigure} \\
\begin{subfigure}{\columnwidth} 
    \includegraphics[width=\textwidth]{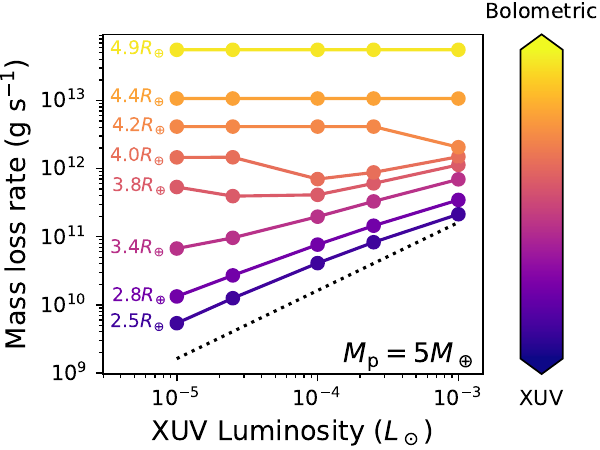}
\end{subfigure} \\
\begin{subfigure}{\columnwidth} 
    \includegraphics[width=\textwidth]{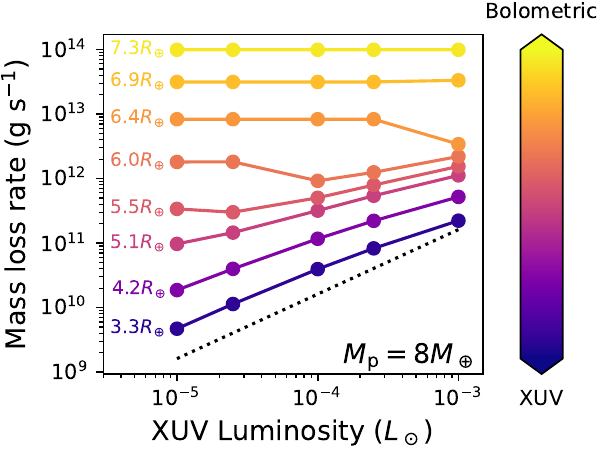}
\end{subfigure}
\caption{Mass loss rates determined from our hydrodynamic modeling for planets with masses of 2, 5, and 8 $M_\oplus$, all with $T_\mathrm{eq} = 1000$~K and $\gamma=1$, as functions of incident XUV fluxes for different planet radii (labeled and represented by colors). At large radii, planets lose mass via a bolometrically driven wind, and there is no dependence on XUV flux. For smaller-radius planets, for which the UV luminosity penetrates the sub-sonic region, the mass loss rate increases with UV flux. However, this dependence is weaker than the $\dot{M} \propto F_\mathrm{XUV}$ slope predicted by Eq.~\ref{eq:energy_limit_traditional} (shown by dotted line for reference). The mass loss rate is not monotonic in UV flux: when the UV barely penetrates the flow, the heating from the ionization does not compensate the decrease in molecular weight.}
\label{fig:UV_vs_Mdot}
\end{figure*}
Regardless of the UV flux, we observe in Fig.~\ref{fig:Rcomp_fixedUV} that the UV absorption radius increases with increasing planet radius, since the absorption occurs at a nearly constant density $\rho \sim 10^{-14}$\ g~cm$^{-3}$. Since $\dot{M}_\mathrm{EL} \propto R_\mathrm{XUV}^2 R_\mathrm{t}$ (Eq.~\ref{eq:energy_limit_traditional}), it is not surprising that increasing the transit and absorption radii increases the atmospheric escape rate. However, the ultimate mass loss rate is not set fully by the traditional energy-limited escape rate. This is made clear in Figure~\ref{fig:UV_vs_Mdot}, which shows the mass loss rate as a function of XUV flux for a range of planetary radii, shown by colors and labels. At large radii (lighter colors), in the bolometric regime, the mass loss rate is independent of XUV flux. At the smallest radii (darker colors), the mass loss rate decreases by a power law with LUV flux. However, the dependence is weaker than $\dot{M} \propto F_\mathrm{XUV}$, the slope predicted by energy limit described in Eq.~\ref{eq:energy_limit_traditional} and shown with a dotted line, indicating that the efficiency of escape decreases at larger fluxes. While the finding of a shallow slope is superficially similar to the predicted TEMP regime \citep{Tang2025a}, we do not find it arises from the same physics: their slope decreases due to an empirical temperature-flux scaling which we do not obtain, while in our models, the temperature, ionization, and mean molecular weight profiles are all changing self-consistently, contributing to the escape rates we calculate. We compare these scalings in more detail in Sec.~\ref{sec:discussion_comparison}. Fig.~\ref{fig:UV_vs_Mdot} demonstrates our ability to predict mass loss rates for a wide range of observationally pertinent sub-Neptune masses, radii, and XUV fluxes.

At intermediate radii, between 3.8 and 4.2 $R_\oplus$ for $M_\mathrm{p} = 5 M_\oplus$ in Fig.~\ref{fig:UV_vs_Mdot}, we observe that the mass loss rate at constant radius can decrease with increasing UV flux. This phenomenon occurs just as the flow is transitioning in fluxspace between UV- and bolometrically-driven escape, a transition which occurs at slightly smaller radii for lower XUV fluxes due to their weaker heating ability. The physical mechanism is similar to that which produces the dip in escape rate below the bolometric rate for fixed UV flux and changing radius shown in Fig.~\ref{fig:Mdot_comparison}: when the UV just penetrates the flow, the mean molecular weight drops, meaning the sonic point is reached at a smaller radius. But most of the density decrease has already occurred in the bolometrically-heated regime, so the additional heating has little impact, and the overall mass loss rate is lower. As we will see in our evolution models in Sec.~\ref{sec:evolution} below, the radius tends to decrease faster than the UV flux does for young puffy planets, so this effect merely slows the decrease in atmospheric escape rate rather than increasing it for a given planet in time.

In summary, these hydrodynamic radiative-transfer simulations capture the transition between core-powered/bolometrically-driven and photoevaporative/energy-limited escape through an intermediate escape regime. We find that atmospheric escape fits analytic estimates at either end of a planet's evolution, but uncover unique and novel physical processes operating in this intermediate regime.

\subsubsection{Effect of atmospheric opacities}\label{sec:results_opacities}
\begin{figure*}
    \centering
    \begin{subfigure}{0.5\textwidth} 
       \includegraphics[width=\textwidth]{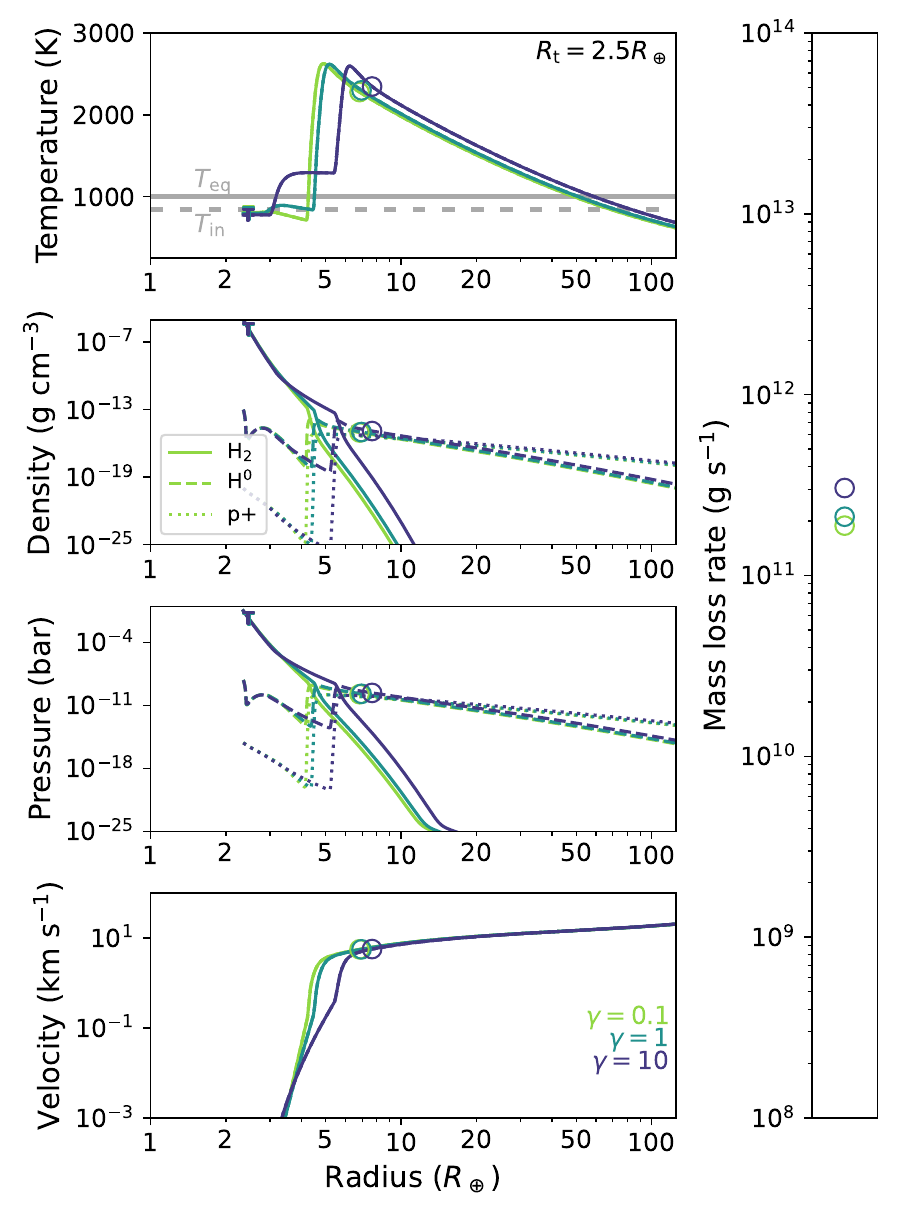}
    \end{subfigure}%
    \begin{subfigure}{0.5\textwidth} 
       \includegraphics[width=\textwidth]{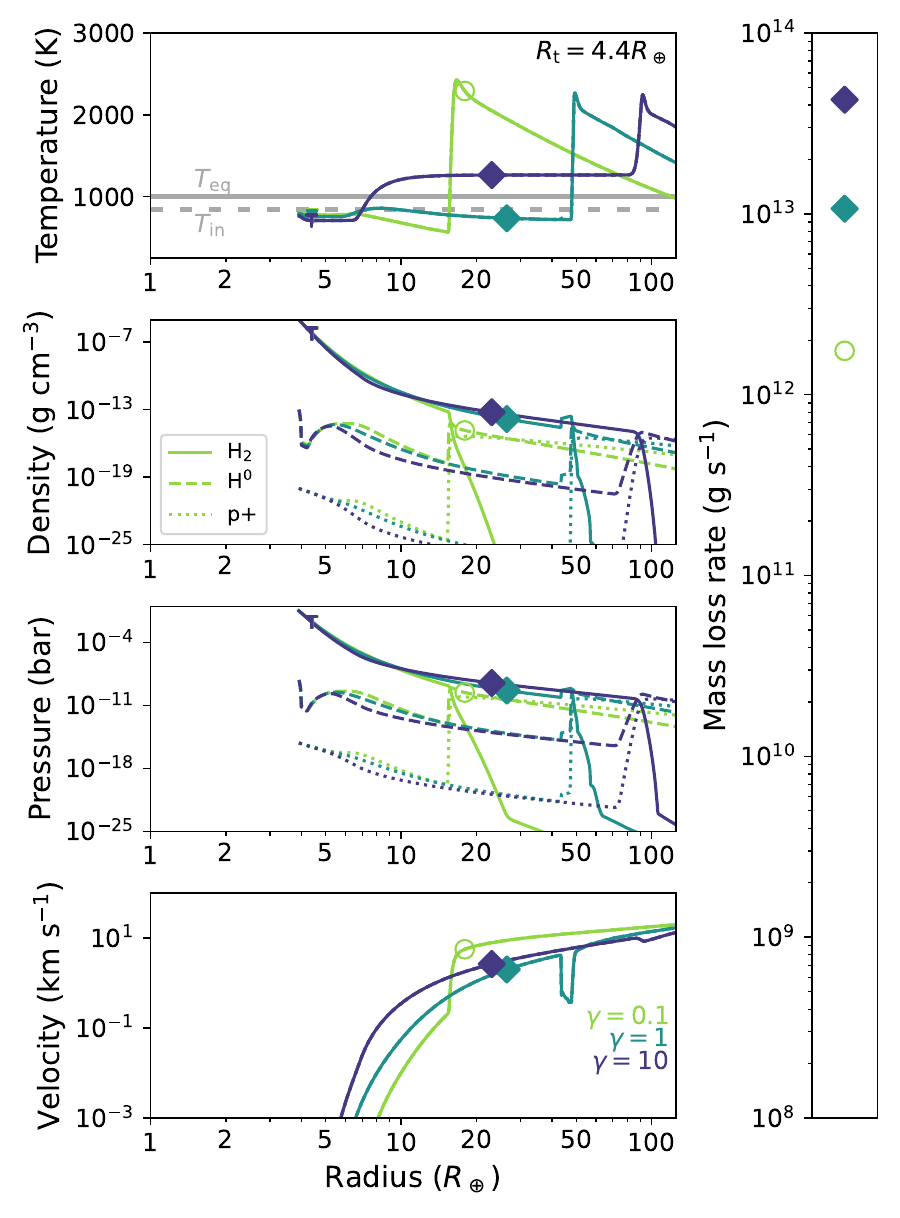}
    \end{subfigure}%
    \caption{Atmospheric profiles for a planet with $M_\mathrm{p} = 5 M_\oplus$, $T_\mathrm{eq} = 1000$~K, and $L_\mathrm{XUV}/L_\mathrm{bol} = 10^{-3}$. Colors represent $\gamma$ values of 0.1, 1, and 10. On the left is $R_\mathrm{t} = 2.5 R_\oplus$, and on the right is $R_\mathrm{t} = 4.4 R_\oplus$. A square denotes the sonic radius of the outflow. A higher $\gamma$ value leads to higher temperatures in the deep radiative region. These hotter temperatures cause a slower drop in atmospheric density with radius, allowing high-energy radiation to penetrate less deep into the atmosphere. For the smaller-radius planet, dominated by photoevaporative loss, this has a small effect on the overall mass-loss rate. For the larger-radius planet, the difference in absorption radius can change whether the UV is absorbed below or above the bolometric sonic point. Therefore, a higher $\gamma$ value leads to a longer-lived core-powered phase before photoevaporation takes over, with higher mass loss rates than predicted in standard CPML \citep{MisenerSchulik2025}.}
    \label{fig:gamma_comp}
\end{figure*}
As shown in \citet{MisenerSchulik2025} and \citet{Schulik2025moon}, atmospheric opacities in general and specifically the ratio between solar and thermal opacities $\gamma$ can strongly influence bolometrically-driven mass loss rates through their effects on the density structure of the atmosphere. Large values of $\gamma$ lead to enhanced escape by increasing the temperature in the bolometric region and therefore the density at the sonic point, while small values conversely inhibit escape by lowering temperatures. To test the influence of atmospheric opacities on our combined photoevaporative-bolometric model, we run models over a range of radii with $\gamma$ values of 0.1, 1, and 10 to capture the range of likely values in sub-Neptunes due to planet-to-planet metallicity and aerosol variations.

In Figure \ref{fig:gamma_comp} we present the profiles obtained in six AIOLOS simulations, all for a $5 M_\oplus$ planet at $T_\mathrm{eq} = 1000$~K and $L_\mathrm{XUV}/L_\mathrm{bol} = 10^{-3}$. On the left are three runs with a transit radius of $2.5 R_\oplus$, while on the right are three runs with a transit radius of $4.4 R_\oplus$. The three colors from light to dark shades represent $\gamma = 0.1$, 1, and 10. The symbols remains the same as in Figure \ref{fig:Rcomp_fixedUV}.

For the smaller-radius planet on the left of Figure \ref{fig:gamma_comp}, we find that the value of $\gamma$ makes less than a factor of two difference in mass loss rate. The $\gamma=10$ simulation is slightly enhanced as its increased density pushes the sonic point slightly outward, but in this case the UV radiation penetrates so deeply that in all three cases the acceleration is dominated by the UV heating, leading to energy-limited escape.

Similar arguments apply to the $\gamma=0.1$ case for the larger-radius planet shown on the right of Figure \ref{fig:gamma_comp}: the flow is driven by UV photoevaporation, although throttled by the long decrease in density at the bolometric scale height. However, we observe different behavior in the $\gamma=1$ and $10$ case. Here, the density falls off more slowly at large radii due to the larger temperature in the bolometrically heated region. This increased density pushes the UV $\tau=1$ surface beyond the sonic point of the purely bolometric flow at $\sim 25 R_\oplus$. Therefore, the $\gamma=1$ and 10 simulations are bolometrically-driven for $R_\mathrm{t} = 4.4 R_\oplus$, while the lower value of $\gamma$ is photoevaporative. Notably, the total escape rate is much higher for the bolometric cases, due to the much larger sonic point and the higher density there, as expected from Equation~\ref{eq:isothermal_loss}. We also note the apparent formation of a shock in the supersonic region of $\gamma=1$ and $\gamma=10$ profiles here, which we discuss further in Appendix~\ref{sec:shock}.

Finally, we remark that an atmosphere with $\gamma \rightarrow\infty$ (i.e. a no-cooling limit) approaches its energy-limited escape rate, no matter the energy source \citep{Schulik2025moon}. Bolometric radiation is, however, absorbed with finite efficiency, quantified by our $\gamma$ terms, at atmospheric radii which overlap with some UV bands. For physical opacities we therefore see photoevaporative escape which follows $\dot{M}\propto F_{XUV}$, but with a decrease from perfect efficiency, as shown in \citet{SchulikBooth2023}.

\subsubsection{Trends with planet mass and equilibrium temperature}\label{sec:results_trends}
\begin{figure*}
    \centering
    \begin{subfigure}{0.5\textwidth} 
       \includegraphics[width=\textwidth]{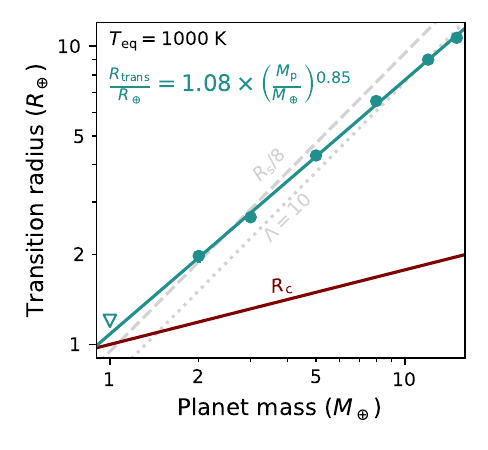}
    \end{subfigure}%
    \begin{subfigure}{0.5\textwidth} 
       \includegraphics[width=\textwidth]{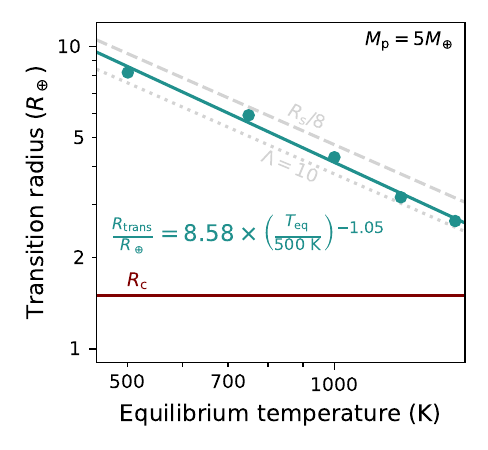}
    \end{subfigure}%
    \caption{The transit radius at which core-powered mass-loss (at large radii) transitions to photoevaporation (at small radii) for different masses (left) and equilibrium temperatures (right) for a XUV flux of $L_\mathrm{XUV} = 10^{-3} L_\mathrm{bol}$ and $\gamma=1$. The gray, dotted lines show constant fractions of the isothermal sonic radius (Equation~\ref{eq:sonic_radius}), and the maroon line shows $R_\mathrm{c}$, the radius of a rocky core with no atmosphere (Equation~\ref{eq:Rcore}). The transition radius increases with increasing mass and decreases with increasing equilibrium temperature, indicating that the lowest mass and hottest planets predominantly undergo core-powered mass loss, while more massive and cooler planets undergo photoevaporation. We also present our power law fits for the transition radii as functions of core mass and equilibrium temperature.}
    \label{fig:transition_radius}
\end{figure*}

To determine the predominant escape regimes over the wide range of observed exoplanet masses and equilibrium temperatures, we run AIOLOS models for planet masses between 1 and 15 $M_\oplus$ and equilibrium temperatures between 500 and 1500~K, each at the three opacity ratio values $\gamma =$ 0.1, 1, and 10 discussed in Section~\ref{sec:results_opacities}. For each combination of planet mass, equilibrium temperature, and $\gamma$, we determine the planet transit radius at which the flow transitions from core-powered to photoevaporative, i.e., the radius at which UV penetrates the bolometric sonic radius by simulating a grid of radii. 
In Figure \ref{fig:transition_radius} we plot the transition radii for $\gamma=1$ and $L_\mathrm{XUV} = 10^{-3} L_\mathrm{bol}$ as functions of planet mass for $T_\mathrm{eq} = 1000$~K (left) and of equilibrium temperature for $M_\mathrm{p} = 5 M_\oplus$ (right). While taking the midpoint between two runs implies some uncertainty in the exact transition radius value, the grid is sufficiently fine that the typical transition radius uncertainty is 5 percent or less, smaller than the size of the points in these plots. For $M_\mathrm{p} = 1 M_\oplus$, we found no run for which the UV penetrates the sonic radius, so we provide an upper limit on the transition radius shown by an unfilled triangle.

In general we find that the transition radii increase with planet mass and decrease with equilibrium temperature, matching the predictions of \citet{OwenSchlichting2024} (see their Figure 6), who predicted that the transition radius should scale linearly with the sonic radius, $R_\mathrm{trans} \approx R_\mathrm{s. iso}/8$, which is shown by the gray dashed line in Figure~\ref{fig:transition_radius}. We find that increase with planet mass is slightly slower than the linear scaling prediction of \citet{OwenSchlichting2024}, with a fit of the $\gamma=1$ results
producing a power of 0.85.
Meanwhile, the trend in equilibrium temperature is closer to linear, though slightly steeper, with a power of $-1.05$ obtained by fitting our data. Combining these linear fits, we obtain a combined fit of the transition radius as a function planet mass and equilibrium temperature:
\begin{equation}\label{eq:transition_fit}
    R_\mathrm{trans} \simeq 4.2 R_\oplus \left(\frac{M_\mathrm{p}}{5M_\oplus}\right)^{0.85} \left(\frac{T_\mathrm{eq}}{1000~\mathrm{K}}\right)^{-1.05} .
\end{equation}

The trends we observe in core mass and equilibrium temperature as shown in Figure \ref{fig:transition_radius} imply that the smallest and hottest planets must contract the most for the UV irradiation to penetrate the sonic radius, and thus to switch from core-powered to photoevaporative escape, although even still the latter is initially in the transitional phase (see Figure \ref{fig:Mdot_comparison}). Meanwhile, more massive and cooler planets are prone to enter the photoevaporative regime at relatively large radii, before much contraction has occurred. Interestingly, these transition radii are fairly close to the transition between boil-off and photoevaporation found in \citet{KubyshkinaFossati2018}, who also predicted a transition at a constant sonic radius-transit radius ratio, which they quantified using the
restricted Jeans parameter $\Lambda = R_\mathrm{s,iso}/R_\mathrm{trans} \sim 10$, shown by the gray dotted line in Fig.~\ref{fig:transition_radius}. To compare these scalings more explicitly, we calculate the transition radius in Eq.~\ref{eq:transition_fit} in terms of $\Lambda$:
\begin{equation}\label{eq:transition_lambda}
    \Lambda_\mathrm{trans} \simeq 9.0 \left(\frac{M_\mathrm{p}}{5M_\oplus}\right)^{0.15} \left(\frac{T_\mathrm{eq}}{1000~\mathrm{K}}\right)^{0.05} .
\end{equation}

However, at lower $\Lambda$, i.e. more extended atmospheres than the transition radii, we find that atmospheric escape is throttled rather than enhanced above the energy limit. \citet{Tang2025b} also find reduced escape for inflated planets, which they explain through UV flux contributing to thermally heating the outflow rather than gravitationally unbinding material, physics our model also includes.
The transition radii we find appear smaller for similar masses and radii than those found in \citet{RogersOwen2024}. We discuss why our results differ from these studies further in Sec.~\ref{sec:discussion}.

\begin{figure*}
    \centering
    \begin{subfigure}{0.5\textwidth} 
       \includegraphics[width=\textwidth]{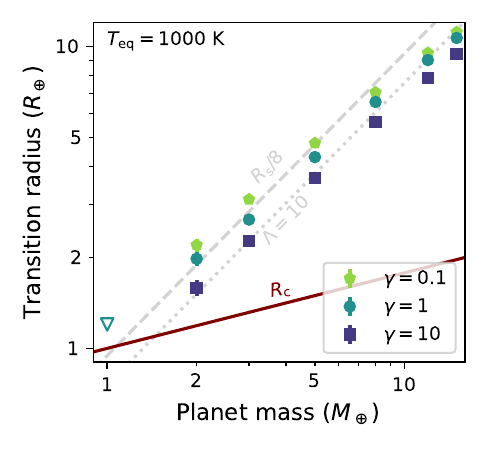}
    \end{subfigure}%
    \begin{subfigure}{0.5\textwidth} 
       \includegraphics[width=\textwidth]{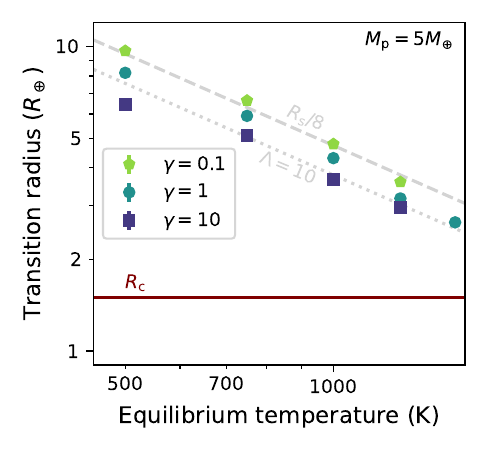}
    \end{subfigure}%
    \caption{The transit radius at which core-powered mass-loss (at large radii) transitions to photoevaporation (at small radii) at different atmospheric $\gamma$ values, for different masses (left) and equilibrium temperatures (right) and a XUV flux of $L_\mathrm{XUV} = 10^{-3} L_\mathrm{bol}$. The transition radius of $\gamma=0.1$ runs are shown with green pentagons, while those of $\gamma=1$ runs are shown with blue circles and those of $\gamma=10$ runs are shown with purple squares. The value of $\gamma$ in the atmosphere can affect the transition radius by $\sim 20$~percent, but the main trends in planet mass and equilibrium temperature are preserved.}
    \label{fig:transition_radius_gammas}
\end{figure*}
We can also test the effects of atmospheric composition on the transition between core-powered and photoevaporative escape. In Figure~\ref{fig:transition_radius_gammas}, three values of the atmospheric opacity parameter $\gamma$ are shown, with the colors matching those of Fig.~\ref{fig:gamma_comp}. The value of the opacity ratio $\gamma$ has a noticeable effect on the transition radius. As described in Section \ref{sec:results_opacities}, a high value of $\gamma$ can push the UV $\tau=1$ surface outward, while conversely a low value leads to deeper absorption (see Figure \ref{fig:gamma_comp}). In other words, a planet with a high value of $\gamma$ needs to contract more to transition to photoevaporation, which will happen at a lower radius, while a planet with a low value of $\gamma$ will transition at a higher radius from core-powered mass loss to photoevaporation. This logic is borne out in the results shown in Figure \ref{fig:transition_radius}: increasing $\gamma$ from 1 to 10 decreases the transition radius, while decreasing $\gamma$ conversely leads to an increased transition radius. The change in transition radius from a factor of 10 change in $\gamma$ ranges between 10 and 20 percent. Changes in $\gamma$ during evolution may be accompanied by increases in the mean molecular weight, which we do not model here. Increasing the mean molecular weight alters both the transit and sonic radii and changes the opacity structure. We discuss these changes further in Sec.~\ref{sec:discussion_evol}.

\section{Time Evolution}\label{sec:evolution}
To model the evolution of small planets in time, we must connect our AIOLOS radiative-transfer hydrodynamic models to interior structure models of sub-Neptunes with H/He-rich envelopes surrounding rocky interiors. To do so, we generally follow the methods described in \citet{MisenerSchulik2025}, which we summarize in Sec.~\ref{sec:evolmethods}. We then present our results for the resolved evolution of planets including boil-off, core-powered, and photoevaporative regimes in Fig.~\ref{sec:evolresults}.

\subsection{Evolution Methods}\label{sec:evolmethods}
\subsubsection{Extrapolation to interior structure}
In short, we extend the base of the AIOLOS domain, which we fix at $\rho=10^{-5}$~g~cm$^{-3}$ such that it is weakly optically thick, to greater depths by approximating the thermal profile with radiative diffusion, such that 
\begin{equation}\label{eq:dlogTdlogP}
    \frac{\partial \ln T}{\partial \ln P} = -\frac{3 \kappa_\mathrm{R} P L}{64 \pi G M_\mathrm{p} \sigma T^4},
\end{equation}
where $R$, $P$, and $T$ are the local radius, pressure, and temperature. $L$ is the planet's luminosity, and $\kappa_\mathrm{R, therm}$ is the local Rosseland mean opacity. The radial pressure gradient in the radiative diffusion region is well-approximated by hydrostatic equilibrium, since the outflow velocities are negligible deep in the envelope:
\begin{equation}\label{eq:dPdr}
    \frac{\partial P}{\partial R} = -\frac{G M_\mathrm{p}}{R^2} \frac{\mu P}{k_\mathrm{B} T}.
\end{equation}
This approximation neglects the effects of advective cooling in the deep envelope, which may be important at early times during the boil-off phase \citep{OwenWu16, RogersOwen2024, TangFortney2024}.

This diffusive, radiative region extends until the temperature gradient, Eq.~\ref{eq:dlogTdlogP}, is equal to the adiabatic gradient, which we take to be 2/7 for the H$_2$ dominated envelopes we consider here. This point is the radiative-convective boundary, $R_\mathrm{rcb}$. From $R_\mathrm{rcb}$, we model the envelope as fully convective and adiabatic to the core radius, $R_\mathrm{c}$, which we approximate using a relation appropriate for an incompressible silicate of mass $M_\mathrm{p}$ \citep{Valencia06}:
\begin{equation}\label{eq:Rcore}
     R_\mathrm{c}/R_\oplus \simeq (M_\mathrm{p} /M_\oplus)^{1/4} .
\end{equation}
This is also in practice the lowest the base of the atmosphere can be in our model.

These atmospheric structures correspond to planet envelope masses and energies that dictate the planetary evolution. The planet envelope mass can be found by integrating the density profile. The total available energy for cooling is the sum of the available energy in the core and atmosphere, $E = E_\mathrm{core} + E_\mathrm{atm}$.  The atmospheric energy is given by
\begin{equation}\label{eq:E_atm}
    E_\mathrm{atm} = \int_{R_\mathrm{c}}^{R_\mathrm{s}} 4 \pi r^2 \left( -\frac{G M_\mathrm{p}}{r} + \frac{1}{\gamma_\mathrm{ad}-1} \frac{k_\mathrm{B} T(r)}{\mu} \right) \rho(r) \mathrm{d}r
\end{equation}
where the adiabatic index $\gamma_\mathrm{ad}=7/5$ for a diatomic atmosphere. Meanwhile, since we take the core to be incompressible, the core's available energy is purely thermal: 
\begin{equation}\label{eq:E_core}
    E_\mathrm{core} = \frac{1}{\gamma_\mathrm{core}-1} \frac{M_\mathrm{p}}{\mu_\mathrm{core} k_\mathrm{B} T_\mathrm{core}}
\end{equation}
Here $\gamma_\mathrm{core} \sim 4/3$ is the core's `adiabatic index', a representation of its specific heat capacity \citep{Scipioni17, MS21}, and $\mu_\mathrm{core} =60$~amu is the core's mean molecular weight. We assume the core is isothermal and thermally coupled to the base of the atmosphere, such that $T_\mathrm{core} = T(r = R_\mathrm{c})$. These integrals then allow us to calculate the atmospheric mass loss and cooling timescales: $t_\mathrm{loss} = M_\mathrm{atm}/\dot{M}$ and $t_\mathrm{cool} = E_\mathrm{atm}/\dot{E}$, respectively.

\subsubsection{Time Evolution Procedure}
We model planetary thermal and mass evolution following the methods of \citet{MS21} and \citet{MisenerSchulik2025}. We begin by specifying an initial planet radius and atmospheric mass and calculating the corresponding luminosity $L$, base of the AIOLOS domain, i.e. $R(\rho=10^{-5}$~g~cm$^{-3}$), and core and atmospheric energies using our structure model. We choose the initial radius following the method of \citet{MisenerSchulik2025} by cooling an envelope with a specified initial atmospheric mass for 10~Myr without any escape from an arbitrarily large radius, producing an envelope with a cooling time $\sim 10\times$ older than its age and thus approximating the boil-off regime \citep{OwenWu16, RogersOwen2024}.

To best conserve computational resources, for each planet of mass $M_\mathrm{p}$, equilibrium temperature $T_\mathrm{eq}$, and atmospheric opacity $\gamma$, we take the common approach \citep[e.g.][]{baraffe2004, LopezFortney2013, ChenRogers2016, MisenerSchulik2025} of running a grid of hydrodynamic models over a range of planet radii and UV luminosities, then interpolating to find the mass loss rate that corresponds to the current $R_\mathrm{p}$ and $L_\mathrm{XUV}$. We interpolate using the using the SCIPY \texttt{interpolate.griddata} method \citep{scipy}, with our grid spacing and bounds varying to sufficiently capture the transitions in mass loss regime for the wide range of planets we study here.

At each timestep, we evolve the planet in time by $\Delta t= 0.01 \min{(t_\mathrm{cool}, t_\mathrm{loss})}$. We find the new atmospheric mass $M_\mathrm{atm, new} = M_\mathrm{atm} - \dot{M} \Delta t$ and planet energy $E_\mathrm{new} = E - L \Delta t -\Delta E_{\dot{M}}$, using the mass loss rate from the hydrodynamic grid interpolation, using $L_\mathrm{XUV}(t)$ following Eq.~\ref{eq:LXUV}, and the planet luminosity from the structure model. The energy loss due to mass loss is the sum of the gravitational and thermal energy of a parcel escaping the planet: $\Delta E_{\dot{M}}= \dot{M} \Delta t (G M_\mathrm{p}/R_\mathrm{s} - c_\mathrm{s}^2)$. We then determine the new atmospheric structure, specified by the luminosity and radius, for the new mass and energy using the \textsc{scipy} \texttt{fsolve} root finding method, with a tolerance of $10^{-8}$ \citep{scipy}. These new atmospheric parameters beget a new interpolated mass loss rate, and we proceed with the iterations.

\subsection{Evolution Results}\label{sec:evolresults}
\begin{figure*}
\hspace*{-0.5cm}
\includegraphics[width=\textwidth]{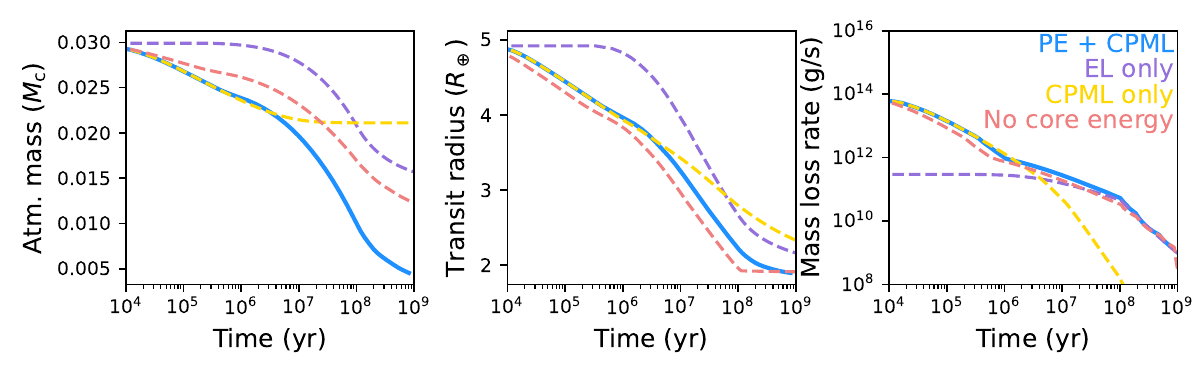}
\caption{Evolution in atmospheric mass (left), transit radius (center), and mass loss rate (right) for three models of a $5 M_\oplus$ planet at $T_\mathrm{eq} = 1000$~K, with $M_\mathrm{atm}(t=0)=0.03 M_\mathrm{p}$ and $R_\mathrm{t}(t=0) = 5.0 R_\oplus$. In blue is our fiducial model including core-powered and photoevaporative escape. For comparison, in red is a model with no core thermal energy included. In this model, the planet contracts more quickly than our standard model and loses much less mass as a result. In yellow is a model including only bolometric escape (CPML-only). Here, photoevaporation never takes over, and the planet also loses less mass than in the combined model. In purple is a model assuming energy-limited photoevaporative escape (Eq.~\ref{eq:energy_limit_traditional}) at all times. This planet loses more mass at early stages and thus contracts faster than the combined model. These results show that the core energy, high energy radiation from the host star, and the transition from bolometric to photoevaporative escape must be considered to accurately capture the evolution of sub-Neptunes.}
\label{fig:evol}
\end{figure*}
In Figure~\ref{fig:evol}, we show the results of our evolution using interpolation of our hydrodynamic results for a planet with $M_\mathrm{p} = 5 M_\oplus$, $T_\mathrm{eq} = 1000$~K, $\gamma=1$, an initial atmospheric mass of 3\% $M_\mathrm{p}$, and an initial transit radius of $5.0 R_\oplus$. In blue we have our fiducial model which interpolates the self-consistent mass loss rates obtained from our hydrodynamical model, and shown in Sec.~\ref{sec:hydroresults}, as a function of planet radius. We see that at early times the escape rate is extremely high, stripping $\sim 20$ percent of the atmospheric mass in $\lesssim 10^6$~yr. The planet also contracts rapidly in this stage. 

This early stage is fully within the bolometrically-driven regime: per Fig.~\ref{fig:transition_radius} and Eq.~\ref{eq:transition_fit}, the transition radius for this planet is $R_\mathrm{trans}=4.2 R_\oplus$. A marked transition occurs when the planet shrinks past this radius, at $\sim 10^6$~yr.
The mass loss rate begins to decrease more slowly with time, producing a kink in the right panel of Fig.~\ref{fig:evol} which indicates the onset of photoevaporation. The higher escape rate leads to efficient stripping over 100~Myr (left panel), and accelerating contraction due to the escape (middle panel). Past 100~Myr, the end of the saturated phase of XUV emission (Eq.~\ref{eq:LXUV}), the escape rate begins to decline steeply due to the drop in XUV luminosity. After 1~Gyr, the planet retains 0.5\% of its mass in hydrogen, losing $>80$\% of its initial supply.

To illustrate the effects of the various mechanisms we consider here, we include three other test cases in Figure~\ref{fig:evol}. In yellow is a `CPML only' case, in which the incident UV flux $L_\mathrm{XUV} = 0$. Thus the mass loss rate follows the yellow dots in Fig.~\ref{fig:Mdot_comparison}. This evolution curve tracks the fiducial blue case at early times, indicating that both models are in the bolometrically-driven regime. However, once the radius is smaller than the transition radius, the mass loss rate first briefly rises above the fiducial case, as seen in Fig.~\ref{fig:Mdot_comparison}, then decreases more quickly with time due to the lack of high-energy radiation. The planet thus loses much less mass over time than in the combined case, ending up with an atmospheric mass fraction of 0.02 and contracting more slowly.

Meanwhile, in purple we show the evolutionary track of a planet following only energy-limited escape using Eq.~\ref{eq:energy_limit_traditional} with $R_\mathrm{XUV}=R_\mathrm{t}$. We find that this planet loses atmospheric mass and contracts much more slowly at early times than the fiducial case, consistent with our findings that the $R_\mathrm{t}^3$ energy-limited escape approximation badly underestimates escape rates (see Fig.~\ref{fig:Mdot_comparison}). Under this mass loss prescription, our standard planet retains 50\% of its initial complement of hydrogen, a larger final atmospheric mass than in our fiducial model.

These results support the importance of a combined treatment of core-powered mass-loss and photoevaporation. As predicted in \citet{OwenSchlichting2024}, planets typically start out undergoing core-powered mass-loss and then transition over time to photoevaporation. As a result, although the observed super-Earth population likely contains significant fractions of planets where each mechanism controlled the final removal of the H/He envelope, photoevaporation is typically responsible for atmospheric escape at late times and hence the final carving of the exoplanet radius valley \citet{OwenSchlichting2024}. Therefore, despite the ability of core-powered mass-loss to remove exoplanet atmosphere over Gyr timescales when modeled in isolation \citep{GuptaSchlichting2021}, it typically precedes photoevaporation in a self-consistent combined model, dominating the mass-loss at early times before giving way to photoevaporative escape. The modeling of the physical interaction between the two mass-loss mechanisms and their evolution in time is therefore crucial in providing predictions for the correct interpretation of the results of current and future exoplanet demographic studies \citep[e.g.][]{BergerHuber2020, DavidContardo2021, ChristiansenZink2023}. 

Finally, in red in Figure~\ref{fig:evol} is a model in which we fully neglect the core's thermal inertia; i.e., we set $E_\mathrm{core}$ (Eq.~\ref{eq:E_core}) to 0 when calculating the energy evolution, such that cooling solely decreases the energy of the envelope. We see that neglecting the core energy also leads to less atmospheric escape, with the atmospheric mass being $\sim 3 \times$ larger than the fiducial case at $t=100$~Myr. This decrease in mass loss rate, which is lower at all times than the fiducial case, is because the envelope is able to contract faster without the core resupplying energy into it (middle panel), the fundamental conceit of the core-powered framework. Therefore, the core's thermal energy also plays a key role in the evolution of small exoplanets, even when the mass-loss is dominated by photoevaporation, and both facets of planet mass and energy balance must be considered for an accurate estimation of planet evolution, as also noted in previous works \citep{LopezFortney14, ChenRogers2016}. In other words, core-powered mass-loss can `enhance' photoevaporation over a significant region of parameter space \citep{OwenSchlichting2024}.

\begin{figure*}
\includegraphics[width=\textwidth]{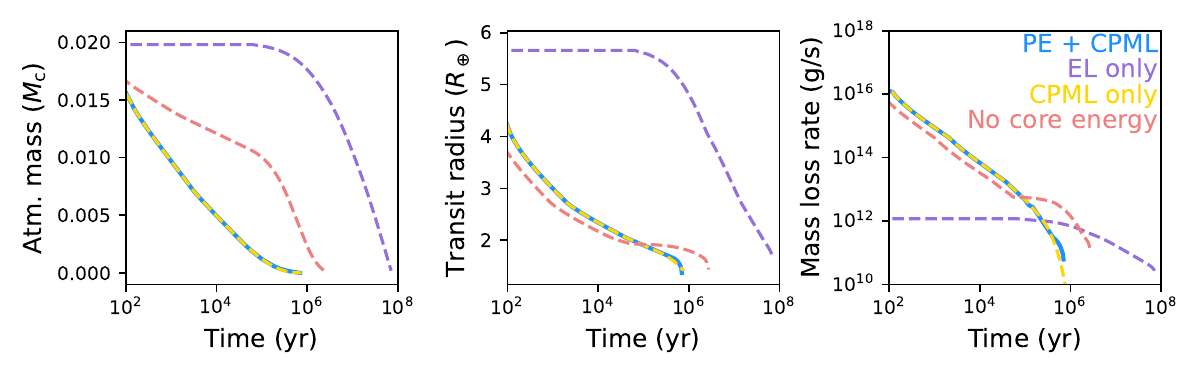}
\caption{Evolution in atmospheric mass (left), transit radius (center), and mass loss rate (right) for three models of a $2 M_\oplus$ planet at $T_\mathrm{eq} = 1000$~K, with $M_\mathrm{atm}(t=0)=0.02M_\mathrm{p}$ and $R_\mathrm{t}(t=0) = 5.8 R_\oplus$. The colors have the same meanings as in Fig.~\ref{fig:evol}. Here atmospheric escape proceeds more quickly from a lower mass planet. The transition radius, at 2.2~$R_\oplus$, is not reached until quite late in the evolution, when most of the envelope has already been stripped. The core energy also plays a vital role in preventing atmospheric contraction. Thus, the atmospheric mass evolution is almost entirely driven by core-powered escape.} 
\label{fig:evol_M2}
\end{figure*}
However, there remain situations where a planet's mass evolution is nearly totally dictated by core-powered escape. We show one such case in Figure~\ref{fig:evol_M2}, for a 2~$M_\oplus$ planet at $T_\mathrm{eq} = 1000$~K, with $M_\mathrm{atm}(t=0)=0.02 M_\mathrm{p}$ and $R_\mathrm{t}(t=0) = 5.8 R_\oplus$. This planet also begins its evolution in the core-powered regime. However, the transition radius for this planet is at 2.2~$R_\oplus$ (see Fig.~\ref{fig:transition_radius}), much smaller than the $5~M_\oplus$ case, which implies that much of the planet's evolution occurs in the core-powered phase. Indeed, we observe this behavior in the evolutionary simulations: the combined (blue) and core-powered (yellow) cases almost completely overlap until very small radii. Nearly all of the atmospheric mass is stripped in the core-powered phase. Energy-limited photoevaporation is a poor approximation of this planet's evolution (purple). We also note that if the core's energy is neglected, the atmosphere contracts quickly, slowing atmospheric escape until after the atmosphere is fully isothermal (red).

\begin{figure*}
\hspace*{-0.5cm}
\includegraphics[width=\textwidth]{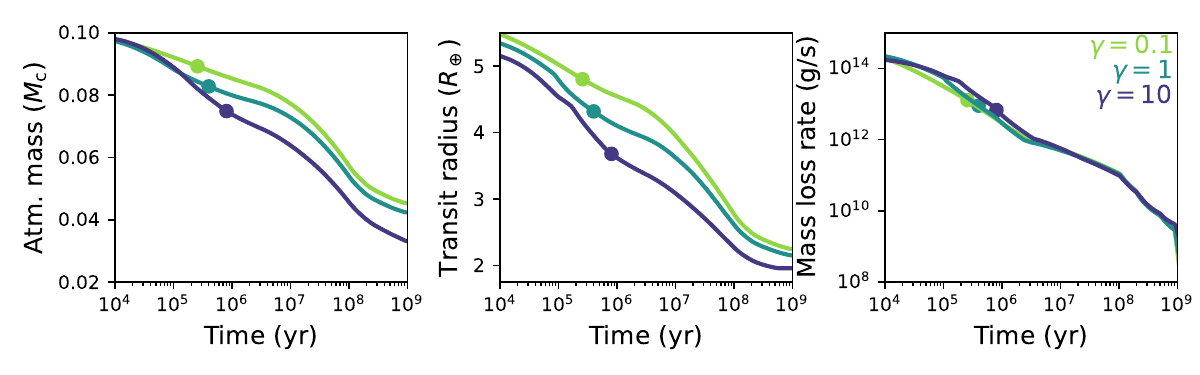}
\caption{Evolution in atmospheric mass (left), transit radius (center), and mass loss rate (right) for three models of a $5 M_\oplus$ planet at $T_\mathrm{eq} = 1000$~K, with $f(t=0)=0.10$ and $R_\mathrm{rcb}(t=0) = 2.5 R_\oplus$, for three different values of the atmospheric opacity ratio $\gamma$. In green is $\gamma=0.1$, $\gamma=1$ is in blue, and $\gamma=10$ in purple. Dots on the curves represent the transition between bolometric- and UV-driven loss, corresponding to the radii denoted in Fig.~\ref{fig:transition_radius_gammas}. Higher values of $\gamma$ lead to enhanced atmospheric escape and more rapid contraction. While the mass loss rates converge, the differences in early evolution lead to differences in evolution on gigayear timescales.} 
\label{fig:evol_gamma}
\end{figure*}
We found in Section~\ref{sec:results_opacities} that the mass loss rates and transition between bolometrically-driven and photoevaporative escape for sub-Neptunes depends on their molecular compositions, through the value of $\gamma$. This implies that the fate of sub-Neptune atmospheres depends on these opacities as well, which we show in Figure~\ref{fig:evol_gamma}. Here we show the time evolution of three models with the same initial conditions but different values of $\gamma$. The $\gamma=1$ case in Fig.~\ref{fig:evol_gamma} is the same as the fiducial case in Fig.~\ref{fig:evol}.

Figure~\ref{fig:evol_gamma} shows that differences in $\gamma$ substantially affect the radial and mass evolution of small planets, even when their escape is driven by photoevaporation. First, the value of $\gamma$ sets the thermal structure of the atmosphere and therefore influences the transit radius and the density at the sonic point \citep{MisenerSchulik2025}. This is evident in the middle panel of Fig.~\ref{fig:evol_gamma}: at the same initial RCB radius, the transit radii are different by $>10$\%. All three planets start in the bolometrically-driven regime, in which the mass loss rates are higher at a given radius for higher $\gamma$ values, although this effect is somewhat offset by their lower radii. Therefore, the low $\gamma$ case loses atmosphere relatively slowly (left panel) and transitions to photoevaporative escape the earliest and at the largest radius, shown by the dots, which represent the radii shown in Fig.~\ref{fig:transition_radius_gammas}. While the mass loss rates converge over time (right panel), the differences in early escape are felt through the rest of the planets' evolution. At all times, the high $\gamma$ model is the smallest in radius, due both to its lower atmospheric mass and its thermal structure. At 1~Gyr, the atmospheric masses are different by a factor of two between the low and high $\gamma$ cases. These results demonstrate the importance of atmospheric composition in assessing the current radii and escape history of sub-Neptunes. Further work will be needed to capture the interactions between outgassing, thermal structure, and mean molecular weight during evolution; we discuss possible directions in Sec.~\ref{sec:discussion_evol}.

\section{Discussion}\label{sec:discussion}
\subsection{Comparison to previous escape models}\label{sec:discussion_comparison}
While this work is not the first to attempt to resolve the transition from a bolometrically-driven to a photoevaporative wind using hydrodynamic models, it encompasses the most relevant physics. In this section, we compare our results to those of selected other works on the subject.

\citet{KubyshkinaFossati2018} was one of the first to study a grid of hydrodynamic models across the sub-Neptune parameter space. They found a sharp change in behavior at a particular ratio between the thermal and gravitational energy of a planet, which they quantify using the restricted Jeans escape parameter $\Lambda$, which is equal to $R_\mathrm{s,iso}/R$ using Eq.~\ref{eq:sonic_radius}. Specifically, they find that transition between what they term ``boil-off'' and photoevaporation happens between $10 < \Lambda< 30$. Our results broadly agree in that we find the transition at $\Lambda\sim 10$ for $\gamma=1$. However, the models of \citet{KubyshkinaFossati2018} did not incorporate a bolometrically-driven hydrodynamic wind,
instead assuming that the UV always penetrates the sub-sonic region (see also \citealt{KrennFossati2021}). We find that this limit is key to understanding the evolution of young, inflated planets.

Our models can be taken to start after the disk has dissipated, but the disk dispersal itself may be vital to early escape. \citet{RogersOwen2024} model boil-off as gradual process by coupling it to the disk dispersal. They use MESA coupled with an isothermal outflow, which transitions from a sub-sonic breeze to a trans-sonic bolometric Parker wind, limited by advection, as the outer density decreases on sub-Myr timescales. They find that a typical sub-Neptune retains atmospheres a few percent of their total mass post boil-off, which they define as ending when the core's luminosity exceeds the change in energy of the atmosphere. Under this definition, they find that planets similar to our fiducial model should undergo a brief `core-powered mass loss' phase, and that at the CPML/PE transition, which they find using a toy model with an instantaneous XUV acceleration, a $5 M_\oplus$, $1000$~K planet should have $M_\mathrm{atm}=0.02 M_\mathrm{p}$ and $R_\mathrm{t} \sim 6 R_\oplus$. These results are fairly similar to, although about $20$~percent larger in radius than, our simulations (cf Fig.~\ref{fig:evol}), so we conclude that this is reasonable agreement. Intuitively, the presence of the disk should moderate mass loss rates, leading to lower rates at the earliest stages compared to our models.

\citet{TangFortney2024} and \citet{Tang2025a} examine the transition between boil-off and core-powered escape.
Building on this work, \citet{Tang2025b} incorporate photoevaporative winds using the hydrodynamic Wind-AE code \citep{BroomeMurray-Clay2025}, proposing a regime,
termed thermal-energy-mediated photoevaporation (TEMP), in which XUV energy is converted into internal energy rather than $P\mathrm{d}V$ work, which
lowers escape rates for large planets and helps preserve the observed super-puff population. 
They propose a scaling $\dot{M} \propto R_\mathrm{XUV}^2 F_\mathrm{XUV}$ once planets reach a certain size, determined by an escape parameter calculated at the sonic point properties, and recover a photoevaporative efficiency of 0.3-0.5 for energy-limited sub-Neptunes, higher than typically assumed. \citet{Tang2025b} model the transition as occurring when the optical depth at the bolometric sonic point becomes less than 1, modeling the envelope structure using an empirical interpolation between an isothermal hydrodynamic and hydrostatic profile.

Our results agree with \citet{Tang2025a} in that we both find that traditional core-powered escape is important predominantly for low-mass, high-temperature planets, while photoevaporation dominates for cooler, higher-mass planets and that modeling the heat from the interior is key to correctly recovering sub-Neptune evolution \citep{LopezFortney14}, leading to core-powered mass-loss, which that work terms `core-enhanced boil off,'
and photoevaporation.
\citet{Tang2025b} also find a transitional regime between bolometrically-driven escape and energy-limited photoevaporation that results in weak scalings of escape rate with UV flux, the `TEMP regime.' However, our physical assumptions and the results we obtain differ from that work. For example, the mass-loss rate scaling obtained by \citet{Tang2025b} relies on an empirical scaling of temperature with UV flux ($T \propto F_\mathrm{UV}^{0.4}$), and assume the molecular weight at the sonic point, while we calculate our thermal and molecular weight profiles self-consistently. We do not obtain the \citet{Tang2025b} temperature scaling, as shown in Fig.~\ref{fig:UVcomp_fixedR}: our temperatures increase more slowly, with $T \propto F_\mathrm{UV}^{0.2}$ in the case shown.
We also find that our flows remain molecular below the XUV absorption radius in the bolometric regime rather than dissociating at $\sim 1 \mu$bar, as found in \citet{Tang2025a}. 
Finally, we find photoevaporation efficiencies more in line with previous work, with values of about 0.05-0.1, which we attribute to  efficient molecular cooling \citep[see also][Fig. 7]{SchulikBooth2023}.
We therefore conclude that our self-consistent radiation-hydrodynamic simulations do not reproduce some aspects of the TEMP regime as described in \citet{Tang2025b}, although this was not the main focus of this study. While beyond our scope here, we expect that further probing the TEMP regime with the AIOLOS code, which includes all the relevant heating and cooling physics by default, would be a fruitful endeavor for future research.

We also highlight the importance of deep envelope thermal profiles in calculation of escape rates. For example,
for a $\gamma$ value close to 0.1, which seems close to that adopted in \citet{TangFortney2024}, we find cooler temperatures in our radiative region in both the skin and outer regions. This appears to arise from that work's adoption of \citet{Guillot2010} profiles, which are calculated at constant gravity. However, as shown in \citet{MisenerSchulik2025}, the difference in radiating and absorbing areas for sub-Neptunes with $\gamma \neq 1$ can lead to different temperature profiles than typically found in gas giants, where the differences in radii between the $\tau=1$ surfaces are negligible. We find here, building on \citet{MisenerSchulik2025}, that the temperatures in these regions are critical for setting the base of the wind and therefore the escape rates and transition radii for sub-Neptunes and thus need careful, self-consistent consideration.

We also find significantly less heating in the XUV regions in our fiducial models, with temperatures 1/5th those attained in \citet{Tang2025b}. Although some of these differences may be due to different opacity assumptions, we have demonstrated that such assumptions have little effect on our derived mass loss rates in Appendix~\ref{sec:appendix_opacity}. We also find a shock when the UV is absorbed in the supersonic region which may prevent relaxation codes such as Wind-AE from recovering isothermal outflow solutions, which we discuss further in Appendix~\ref{sec:shock}.

\subsection{Interpretation of observed escape from small planets}
By synthesizing the two competing mechanisms for atmospheric escape, core-powered and photoevaporative mass loss, this work contextualizes observations of escape from small exoplanets \citep[e.g.][]{ZhangKnutson2023, LoydSchreyer2025, AhrerRadica2025}. We find that most sub-Neptunes likely undergo a phase of core-powered mass-loss which later transitions into photoevaporation. The current regime is determined by the transition radius that depends weakly on the LUV luminosity and strongly on the planet's distance from the star, planet mass, and atmospheric composition through $\gamma$ (Fig.~\ref{fig:transition_radius}), unless the UV luminosity drops so low as to be inefficient compared to bolometrically-driven escape. Wind speeds at the sonic point are faster in photoevaporative flows than in bolometric outflows, due to their higher temperatures and lower mean molecular weights from dissociation and ionization (Fig.~\ref{fig:Rcomp_fixedUV}).  We therefore predict that planets above the transition radii should have slow winds consistent with bolometric escape, while those below the transition radii should have fast, UV-powered winds. We note that a relationship between velocity and mass loss rate is not straightforward, due to the changing molecular weights: in our simulations, outflows are fully molecular in the bolometrically-driven phase \citep[c.f.][]{Tang2025a} and atomic/ionized in the photoevaporative phase. Thus, a fast photoevaporative outflow can correspond to a lower mass loss rate than a slow, core-powered outflow of the same planet size, as we see in Fig.~\ref{fig:UV_vs_Mdot}. 
Measurements of the sound speed at the sonic point, achievable by Ly~$\alpha$ tail length measurements \citep{SchreyerOwen2024}, can constrain both the temperature and molecular weight of planetary outflows and thus diagnose their current escape regime.
We also highlight that in the early stages of planetary evolution, atmospheric escape rates and velocities are highly sensitive to the value of $\gamma$ (Fig.~\ref{fig:gamma_comp}), as discussed in \citet{MisenerSchulik2025}, thus allowing one to probe the atmospheric composition of young planets with escape measurements.

\subsection{Production and composition of super-Earth atmospheres}\label{sec:discussion_evol}
The transition from sub-Neptunes with primary atmospheres to super-Earths with secondary atmospheres has been the subject of much work in the past few years \citep[e.g.][]{KiteBarnett2020, MS21, Krissansen-TottonWogan2024, HengOwen2025, CherubimWordsworth2025}. By modeling the mass evolution of small planets, these results bear implications for these works. In agreement with \citet{OwenSchlichting2024}, planets typically start their evolution after boil-off/spontaneous mass-loss by undergoing core-powered mass-loss and then transition over time to photoevaporation. As a result, photoevaporation is often responsible for atmospheric escape at late times and hence the final carving of the exoplanet radius valley, except for the smallest and hottest planets.

Of particular importance is the evolution of the bulk atmospheric composition, which we have neglected in this work. The composition of the atmosphere has multiple effects: increasing the mean molecular weight decreases the scale height and increases the sonic radius of the outflow, making it more difficult to unbind any remaining atmosphere. Changing the composition also changes the opacity across wavelengths: atomic metals are good X-ray absorbers, changing the XUV absorption radius. In an often neglected aspect, the composition of the atmosphere affects escape through the relative abundance of optical and IR absorbers, quantified by the radiative $\gamma$: many carbon- and oxygen-bearing molecules expected in atmospheres scaled from solar metallicity \citep[e.g.][]{GaoPiette2023} and through atmosphere-interior interactions \citep[e.g.][]{WerlenDorn2025} are good infrared absorbers, leading to a low $\gamma$ value more typical of scaled solar metallicities \citep[e.g.][]{Tang2025a}. These atmospheres will have lower escape rates than our fiducial $\gamma=1$ models. Conversely, silicates and other species expected from mantle outgassing \citep[e.g.][]{MisenerSchlichting2022, MisenerSchlichting2023, PietteGao2023, FalcoTremblin2024}, or accretion of typical primordial species \citep{Schulik2025moon} as well as aerosols, are often good visible absorbers, leading to high $\gamma$ values, thermal inversions, and increased escape rates. These changes in atmospheric composition imply differences in cooling efficiency: radiative diffusion will be less efficient, leading to longer cooling timescales \citep[e.g.][]{Tang2025a}. Different atmospheric compositions also imply different fractionation efficiencies at late times, which may be key to late-stage sub-Neptune evolution \citep{CherubimWordsworth2025}. Any future models capturing this transition with realistic atmospheric evolution and interior coupling will also need to account for changes in deep envelope structure implied by magma-hydrogen interactions, such as molecular weight gradients \citep[e.g.][]{MisenerSchlichting2022, Markham22} and chemical equilibration \citep[e.g][]{SY22,MisenerSchlichting2023, RogersSchlichting2024}.

\section{Conclusion}\label{sec:conclusion}
In this work we have used radiation-hydrodynamic simulations coupled with interior models to calculate atmospheric escape by core-powered mass-loss and photoevaporation self-consistently for the first time. We find that these planets are sculpted by a range of energy sources working to unbind their envelopes, including bolometric and high-energy radiation from their host stars as well as heat released from their cooling interiors. We find that all three sources of energy must be carefully accounted for to produce an accurate picture of sub-Neptune mass evolution and final atmospheric states. Our hydrodynamic modeling confirms the analytic calculations of \citet{OwenSchlichting2024} that the location of the XUV absorption radius vis-a-vis the bolometric sonic point determines the escape behavior of the planet: if XUV radiation is unable to penetrate the bolometrically-driven outflow before it reaches super-sonic velocities, the mass loss rate is unaffected by the XUV instellation and instead depends sensitively on the density profile of the planet. Once XUV photons are able to penetrate well within the putative bolometric sonic point, their heating drives a photoevaporative wind that at small radii is well-approximated by the traditional energy limit, with an efficiency on the order of 10\%. When the XUV radiation is just able to penetrate, the flow is already near sonic, and the XUV heating works mainly to dissociate and ionize the hydrogen, leading to lower efficiencies than expected from pure energy-limited escape. This novel and unanticipated behavior underscores the need for hydrodynamic modeling which can incorporate all the relevant facets of sub-Neptune evolution.

These models have many implications that can be tested in the observed exoplanet population:
\begin{itemize}
    \item We find that core-powered, bolometrically-driven escape dominates when planets have radii larger than a transition radius, such that UV absorption occurs above the bolometric sonic radius. Specifically, the transition radius between the two mass-loss regimes is $\approx 4.2 R_\oplus$ for a $5 M_\oplus$ planet at $T_\mathrm{eq} =1000$~K. The transition radius scales nearly linearly with planet mass and the inverse of equilibrium temperature, as given by Eq.~\ref{eq:transition_fit}. Therefore, the core-powered regime extends to smaller planetary radii at lower masses and higher equilibrium temperatures. The outflows in the photoevaporative regime are hotter and have higher wind velocities than those in the core-powered regime, which is testable by escape observations.
    \item Using coupled escape-evolution models, we find that core-powered mass-loss precedes photoevaporation. Specifically, our combined model shows that a typical sub-Neptune, after boil-off/spontaneous mass-loss, starts out in the core-powered mass-loss regime and then transitions into photoevaporation over time. This is a notable inversion in the predicted timescales of the mass-loss mechanisms in isolation, and observational population-level mass-loss results should be re-evaluated in light of these new results \citep{BergerHuber2020, DavidContardo2021, ChristiansenZink2023}.
    \item We also find that synthesizing the core-powered and photoevaporative models leads to more escape over a typical small planet's lifetime than considering either alone. These phases of atmospheric escape should be considered when extrapolating current atmospheric masses to natal accretion \citep[e.g.][]{RogersOwen21}.
\end{itemize}

Our work also reveals the strong dependence of atmospheric escape on the relative strength of the infrared to optical opacities, and hence the composition of the atmospheres. The presence of infrared absorbers ($\gamma<1$) decreases temperatures in the radiative region, throttling the escape rates at early times, increasing the planetary radius at which the flow transitions to being photoevaporative, and altering the long-term radial and mass evolution of the planet. Conversely, increasing these temperatures, i.e. creating a thermal inversion, through the presence of visible absorbers in the upper atmosphere ($\gamma>1$) promotes more vigorous escape. These results point to the need for coupled atmospheric chemistry and escape codes which can resolve sub-Neptune evolution in composition, mass, and energy, fully linking their atmospheres and interiors, so that we can understand the delicate links between them. Self-consistently coupled models will be necessary to interpret the continuing influx of JWST observations of sub-Neptune atmospheres and escape and to unveil what atmospheric compositions we should expect in their terrestrial super-Earth cousins.

\section*{Acknowledgments}
We thank the anonymous reviewer for their thoughtful feedback on this manuscript. W.M. was supported by the AEThER program, funded in part by the Alfred P. Sloan Foundation under grant \#G202114194. H.E.S. gratefully acknowledges support from the NASA Exoplanet Research Program under grant number 22-XRP22-0118. M.S. and J.E.O were funded by the European Research Council (ERC) under the European Union’s Horizon 2020 research and innovation program (Grant agreement No. 853022, PEVAP). J.E.O. is supported by a Royal Society University Research Fellowship.
\software{NUMPY \citep{numpy},
          MATPLOTLIB \citep{Matplotlib},
          SCIPY \citep{scipy}},
          AIOLOS \citep{SchulikBooth2023, aiolos_zenodo}

\bibliography{cpmlpe} 
\bibliographystyle{aasjournal}

\appendix
\section{Opacity sensitivity testing}\label{sec:appendix_opacity}
\begin{figure}
\hspace*{-0.5cm}
\centering
\includegraphics[width=0.5\columnwidth]{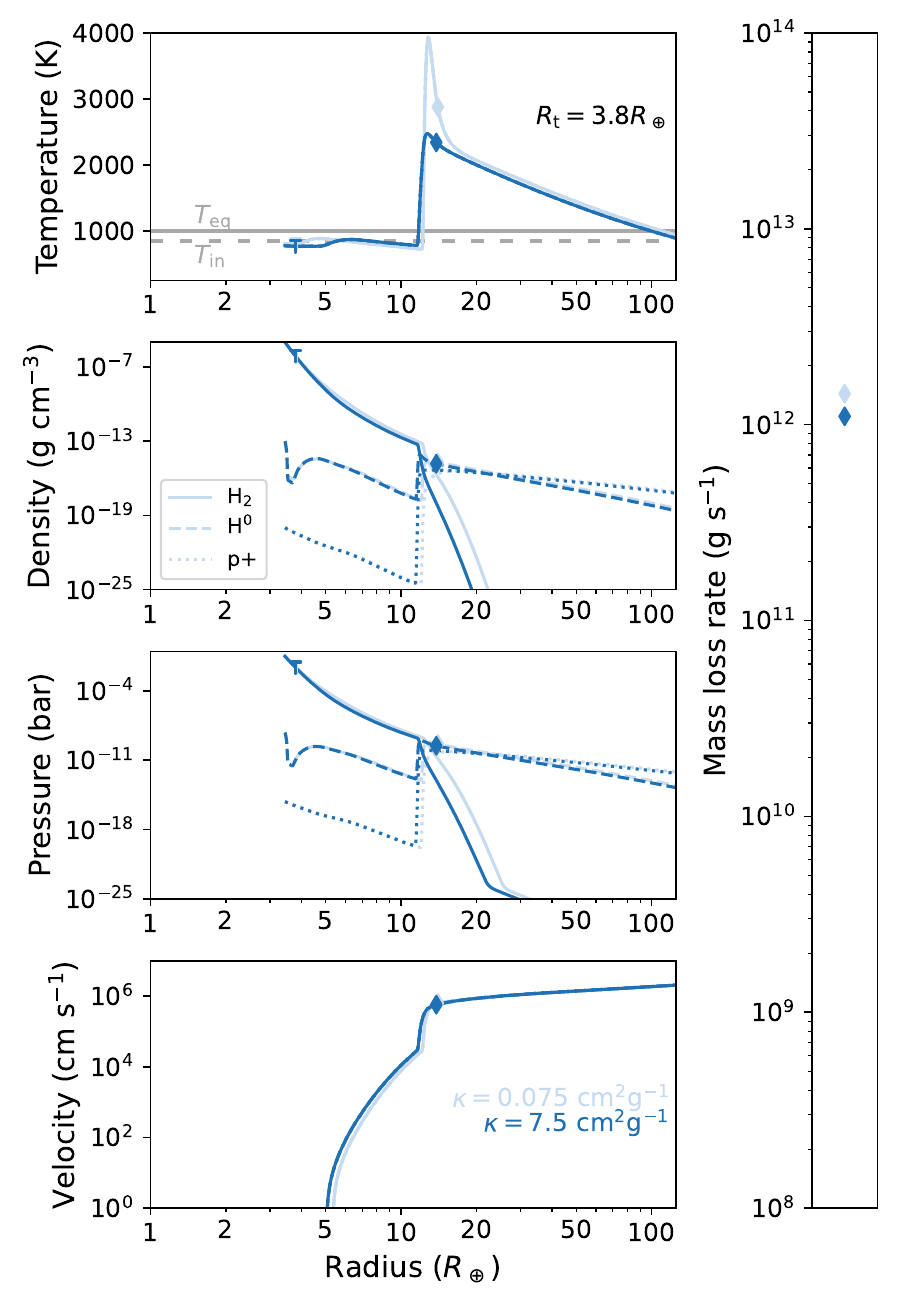}
\caption{Atmospheric profiles for $\gamma \sim 1$, $M_\mathrm{p} = 5 M_\oplus$, $T_\mathrm{eq} = 1000$~K, and $R_\mathrm{t} = 3.8 R_\oplus$ for two different opacities: $\kappa_\mathrm{P, therm} = \kappa_\mathrm{P, \odot} = 7.5$~cm$^2$g$^{-1}$ and 0.075~cm$^2$g$^{-1}$. A symbol denotes the sonic radius of the outflow. The profiles are similar, with the lower opacity atmosphere reaching a higher peak temperature where the UV is absorbed, leading to a $\sim 50$\% increase in the mass loss rate.}
\label{fig:opa_comp}
\end{figure}
In Figure \ref{fig:opa_comp} we test the sensitivity of our results to our opacity assumptions by decreasing the solar and thermal Planck opacities by a factor of 100 (light blue), compared to the standard case (dark blue), identical to the brown curve on the right panel of Figure \ref{fig:Rcomp_fixedUV}. We find that the mass loss rates are nearly identical to less than a factor of two. The lower opacity simulation has decreased cooling efficiency, leading to a larger spike in temperature than the higher opacity simulation and a slightly higher temperature at the sonic point. However, both receive the same UV energy and follow the same bolometric profile, leading to similar mass loss rates.

\section{Shock behavior in supersonic region}\label{sec:shock}
\begin{figure}
    \centering
    \includegraphics[width=\linewidth]{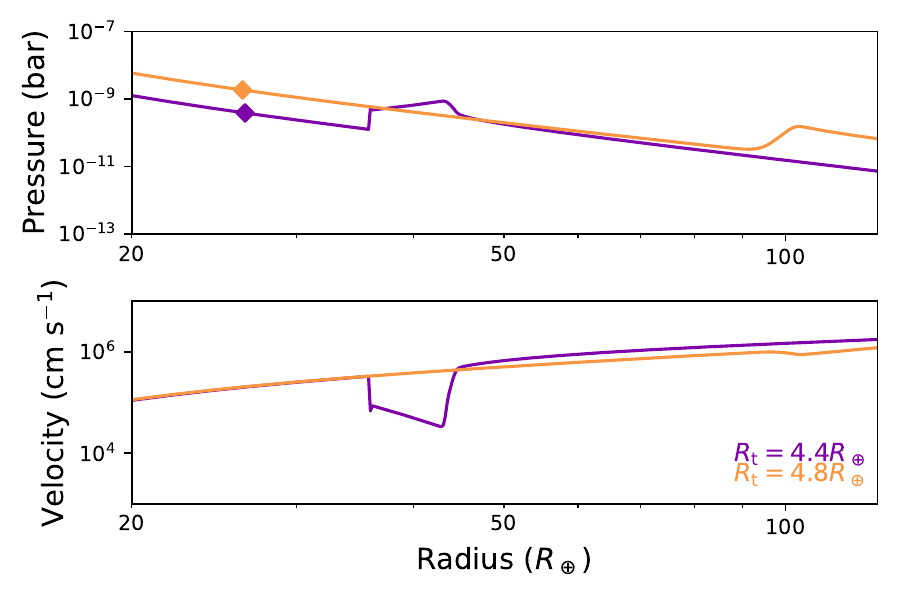}
    \caption{Portions of the total pressure (top) and velocity (bottom) profiles for a $M_\mathrm{p} = 5 M_\oplus$, $T_\mathrm{eq} = 1000$~K planet with $L_\mathrm{XUV} = 10^{-3} L_\odot$ for radius $R_\mathrm{t} = 4.4 R_\oplus$ (purple, the same as the  teal run shown in Figure~\ref{fig:gamma_comp}) and $R_\mathrm{t} = 4.8 R_\oplus$ (orange, the same as the orange run on the right side of Figure~\ref{fig:Rcomp_fixedUV}). The sonic points are shown by diamonds. At the UV absorption front, which is beyond the isothermal sonic point, both these profiles exhibit an increase in pressure and a decrease in velocity. These attributes are consistent with an R-type shock.}
    \label{fig:shock}
\end{figure}
Here we further examine the shock behavior we observe in outflows where the UV absorption front is in the supersonic region. In Figure~\ref{fig:shock} we zoom in on the total pressure and velocity profiles of two aiolos outputs already shown in the main body, in Figures~\ref{fig:Rcomp_fixedUV} and \ref{fig:gamma_comp}. In both these profiles, the flow is already supersonic when the UV is absorbed, at $\sim 30 R_\oplus$ for $R_\mathrm{t}=4.4 R_\oplus$ and at $\sim 100 R_\mathrm{t}$ for $R_\mathrm{t}=4.8 R_\oplus$. At this point, both profiles exhibit an increase in pressure by about an order of magnitude, and a corresponding decrease in velocity. These are both characteristic of a rarefied (R-type) shock occurring at the dissociation/ionization boundary \citep{Kahn1954, Shu1992}. The presence of such a shock may inhibit relaxation-type hydrodynamic codes \citep[e.g.][]{BroomeMurray-Clay2025, Tang2025b} from converging on isothermal outflow solutions, underscoring the necessity of a well-balancing code that can accommodate shocks like AIOLOS to investigate these problems.

\begin{figure}
    \centering
    \includegraphics[width=0.5\linewidth]{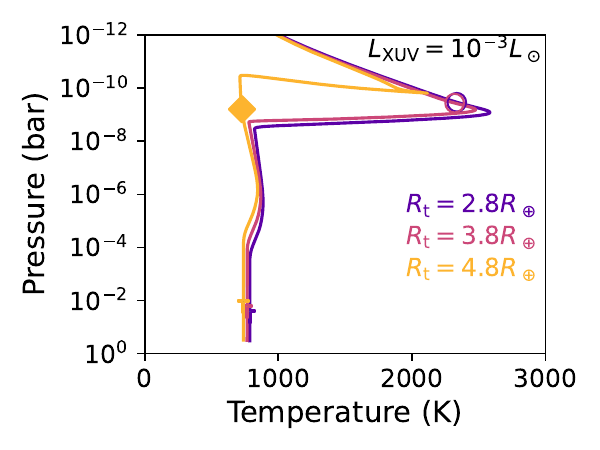}
    \caption{Pressure-temperature version of Figure~\ref{fig:Rcomp_fixedUV}.}
    \label{fig:PT}
\end{figure}

This increase in pressure across the shock is also shown in Figure~\ref{fig:PT}, which shows a pressure-temperature version of the runs shown in Fig.~\ref{fig:Rcomp_fixedUV}. We hope that providing the profiles on these axes as well as those in Fig.~\ref{fig:Rcomp_fixedUV} will aid in comprehension for members of the community who work more often in $P$-$T$ space.

\end{document}